\documentclass[12pt]{article}
\usepackage{amsfonts}
\usepackage{amssymb}
\usepackage{color}
\usepackage{graphicx}

\newcommand{\A}{{\mathfrak A}}

\newcommand{\R}{{\cal R}}
\newcommand{\F}{{\cal F}}
\newcommand{\Sc}{{\cal S}}
\newcommand{\Pc}{{\cal P}}

\newcommand{\mc}{\mathcal}

\newcommand{\be}{\begin{equation}}
\newcommand{\en}{\end{equation}}
\newcommand{\bea}{\begin{eqnarray}}
\newcommand{\ena}{\end{eqnarray}}
\newcommand{\beano}{\begin{eqnarray*}}
\newcommand{\enano}{\end{eqnarray*}}

\newcommand{\1}{1 \!\! 1}
\newcommand{\ST}{\mc S}

\newcommand{\Hil}{\mc H}

\catcode `\@=11 \@addtoreset{equation}{section}
\def\theequation{\arabic{section}.\arabic{equation}}
\catcode `\@=12
\begin{document}

\begin{center}
{\Large \textbf{An operator view on alliances in politics}} \vspace{2cm%
}\\[0pt]

{\large F. Bagarello}
\vspace{3mm}\\[0pt]
DEIM, Facolt\`{a} di Ingegneria,\\[0pt]
Universit\`{a} di Palermo, I - 90128 Palermo, Italy\\[0pt]
E-mail: fabio.bagarello@unipa.it\\[0pt]
home page: www.unipa.it/fabio.bagarello \vspace{8mm}\\[0pt]
\end{center}

\vspace*{2cm}

\begin{abstract}
\noindent We introduce the concept of an {\em operator decision making technique} and apply it to a concrete political problem: should a given political party form a coalition or not? We focus on the situation of three political parties, and divide the electorate into four groups: partisan supporters of each party and a group of undecided voters. We consider party-party interactions of two forms: shared or differing alliance attitudes. Our main results consist of time-dependent decision functions for each of the three parties, and their asymptotic values, i.e., their final decisions on whether or not to form a coalition.
\end{abstract}

\vfill

\newpage


\section{Introduction}

In the last few years the scientific literature has seen a growing interest in the possibility of using quantum ideas and quantum tools in the description of some aspects of several macroscopic systems, systems which, in the common understanding, are usually thought to be {\em purely classical}. This interest has touched very different fields of science, from finance to ecology, from psychology to decision making, and so on. The literature is now very rich, and it increases almost every day. We just cite here some recent monographs, which cover some of the areas just mentioned, but not only: \cite{baa}-\cite{buse}.

Recently, \cite{bagqb}, we have proposed a dynamical approach for a very simple and well known problem in decision making, somehow following what was previously done in \cite{khren2,khren2b}. That work was about a certain version of the two prisoners game, and the main output of our treatment was the deduction of their final decisions (i.e., their decisions when $t\rightarrow\infty$), both when they do mutually interact and when they do not.

Here we apply the same general framework to a problem which is rather common in politics. Let us start with a short historical introduction, useful to motivate our interest in this problem: in 2013, the results of the Italian political elections produced no real winner. Three parties took more or less the same number of votes: the {\em Partito Democratico} (PD), who was really the first party in that election, but which was not strong enough to govern alone, the {\em  Popolo della Libert\`a} (PdL) and the {\em Movimento 5 Stelle} (M5S), both sufficiently strong to have, if allied with PD, much more than the $50\%$ of votes. The other parties took only few votes to really have any power. It was clear to any observer that, to govern, PD had to form a coalition with PdL or with M5S, and the natural choice was to try to form a somehow left-oriented government PD-M5S. However, M5S decided not to ally with PD, despite of the fact that many of its electors disapproved that choice. After a long impasse, Enrico Letta (PD) started to collaborate with PdL and they created a new {\em grand coalition} government, even if a large part of the PD electors was contrary to such a coalition with Berlusconi's party. Summarizing, in a single election we had, in just few weeks, two  similar {\em coalition problems}: should the M5S form a coalition with PD?  And soon after: should the PD form a coalition with PdL?

In many of its decisions M5S asks for some feedback from its supporters, via internet: all the decisions are taken after some poll organized on the web. In particular, the decision on whether to ally or not with PD was the result of an interaction of the Movement with the {\em environment} of its supporters. Moreover, even if  M5S took its decision  essentially after this poll, some of its members also interacted with some of the PD members, looking for some agreement. Of course, in real life, this is not  the end yet: each party also interacts  with other people, the unsatisfied electors of the other parties, or those who usually do not go to vote, the undecided voters, and so on, and also this interaction plays a role in {\em constructing} the final decision. Of course, this does not happen only in Italy. Indeed, this is really rather general and something similar also happened quite recently, in 2014, in France, where the {\em Front National} won in several cities. Its leader claimed she was not going to form any coalition with any other party: was this also what her electors really wanted? Just to mention another example, a similar situation also occured in  Germany, after the  elections in 2005.

Therefore, having clear in mind that we are describing a rather common situation, we focus now on the Italian 2013 election. This is useful to fix some ideas, like for instance the number of the main actors of the model. Our effort will be to derive a dynamics for our system, and to deduce, out of this dynamics, the final decision of the three parties: should they try  to form a coalition or not? And, which are the reasons bringing to this decision? Once a satisfactory model is proposed, is it possible to tune the parameters of the model in order to get different decisions? The general settings we are going to use is the one widely described in \cite{bagbook}, and recently used in \cite{bagqb} for a similar, but simpler, decision making problem. As we will see, we will use three different modes of fermionic operators to model the parties, or, more precisely, their {\em decision functions}, see below, and more modes of fermionic operators, labeled also by a continuous index, to describe the electors of the different parties and the undecided voters. The reason for this particular choice will be discussed in Section II.

Before starting, we like to mention that, to our knowledge, ours is not the first attempt to use mathematics and quantum mechanical tools to model politics. Something in this direction can be found, for instance, in \cite{havkhre} and in references therein. On the other hand, as far as we know, there have been not many other attempts to construct mathematical models in politics, except some old paper like, for instance, \cite{otto}. Slightly more extended is the literature concerning the use of quantum ideas in decision making processes, see e.g. \cite{marti}-\cite{lamb}. In none of these papers, however, an {\em environment} is introduced to drive the dynamical behavior of the system, which, on the other hand, is probably the crucial ingredient of our approach.

\vspace{2mm}

The paper is organized as follows: in  section II we introduce the model and we derive its dynamics. Some particular cases are discussed in Section III, while a more general situation is considered in Section IV. Our conclusions are given in Section V. To keep the paper self contained we discuss some essential (and, for the expert, well known) facts in quantum mechanics in the Appendix.

\section{The model and its dynamics}

In this section we will discuss the details of our model, constructing first the vectors of the players, then the Hamiltonian of the system, and deducing, out of it, the differential equations of motion and their solutions, focusing in particular to their asymptotic (in time) behavior.

In our model we have three parties, $\Pc_1$, $\Pc_2$ and $\Pc_3$, which, together, form the system $\Sc_\Pc$. Each party has to make a choice, and it can choose only one or zero, corresponding respectively to {\em form a coalition} or not. This is, in fact, the only aspect of the parties we are interested in, here. Hence we have eight different possibilities, which we associate to eight different and mutually orthogonal vectors in an eight-dimensional Hilbert space $\Hil_\Pc$. These vectors are called $\varphi_{i,k,l}$, with $i,k,l=0,1$. The three subscripts refers to whether or not the three parties of the models wants to form a coalition at time $t=0$. Hence, for example, the vector $\varphi_{0,0,0}$, describes the fact that, at $t=0$, no party wants to ally with the other parties. Of course, this attitude can change during the time evolution, and deducing these changes is, in fact, the essence of the paper. Analogously, for instance, $\varphi_{0,1,0}$, describes the fact that, at $t=0$, $\Pc_1$ and $\Pc_3$ don't want to form any coalition, while $\Pc_2$ does. The set $\F_\varphi=\{\varphi_{i,k,l},\,i,k,l=0,1\}$ is an orthonormal basis for $\Hil_\Pc$. A generic vector of $\Sc_\Pc$, for $t=0$, is a linear combination of the form
\be
\Psi=\sum_{i,k,l=0}^1\alpha_{i,k,l}\varphi_{i,k,l},
\label{24}\en
where we assume $\sum_{i,k,l=0}^1|\alpha_{i,k,l}|^2=1$ in order to normalize the total probability, \cite{khren2}. In particular, for instance, $|\alpha_{0,0,0}|^2$ represents the probability that $\Sc_\Pc$ is, at $t=0$, in a state $\varphi_{0,0,0}$, i.e. that  $\Pc_1$, $\Pc_2$ and $\Pc_3$ have chosen 0 (no coalition).  As already stated, this framework is quite close to that already introduced in other papers, see \cite{khren2,khren2b}, where certain vectors, in a suitable Hilbert space, are used to describe few essential aspects of the model under consideration. Similar tools (Hilbert spaces, vectors, operators,...) are also used in \cite{marti}-\cite{lamb}, with the main  difference that in these papers the Hilbert spaces are quite often finite-dimensional. This will not be the case for us, because of the role of the electors, which are naturally described using an infinite-dimensional Hilbert space, see below. We believe that this is not just a mathematical trick, but it is also important for the interpretation of the model.

As in \cite{bagqb}, we construct the vectors $\varphi_{i,k,l}$ in a very special way, starting with the vacuum of three fermionic operators, $p_1$, $p_2$ and $p_3$, i.e. three operators which, together with their adjoint, satisfy the canonical anticommutation relation (CAR) $\{p_k,p_l^\dagger\}=\delta_{k,l}$ and $\{p_k,p_l\}=0$. More in details,  $\varphi_{0,0,0}$ is a vector satisfying $p_j\varphi_{0,0,0}=0$, $j=1,2,3$. Of course, such a non zero vector always exists, and it is very useful since it turns out to be an eigenstate of some of the operators used, within our scheme, for the description of the system. Other eigenvectors of these same operators can be constructed
 out of $\varphi_{0,0,0}$:
$$
\varphi_{1,0,0}=p_1^\dagger\varphi_{0,0,0}, \quad \varphi_{0,1,0}=p_2^\dagger\varphi_{0,0,0}, \quad \varphi_{1,1,0}=p_1^\dagger\,p_2^\dagger\varphi_{0,0,0},\quad \varphi_{1,1,1}=p_1^\dagger\,p_2^\dagger\,p_3^\dagger\varphi_{0,0,0},
$$
and so on. Let now $\hat P_j=p_j^\dagger p_j$ be the so-called {\em number operator} of the $j$-th party, which is constructed using $p_j$ and its adjoint, $p_j^\dagger$. Since $\hat P_j\varphi_{n_1,n_2,n_3}=n_j\varphi_{n_1,n_2,n_3}$, for $j=1,2,3$, the eigenvalues of these operators, zero and one, correspond to the only possible choices of the three parties at $t=0$. This is, in fact, the main reason why we have used here the fermionic operators $p_j$: they automatically produce only these eigenvalues\footnote{From this perspective, fermionic operators are to be preferred to other choices, for instance to bosonic operators, since for them the eigenvalues of the number operators are all the natural numbers. Too many!}. Our main effort here consists in {\em giving a dynamics} to these eigenvalues, or, better to say, to the number operators $\hat P_j$ themselves, following the scheme described in \cite{bagbook}. In fact, in this way, we can follow how the parties change their decision with respect to time, regarding alliances. Hence, inspired by a mechanical scheme, it is natural to look for an Hamiltonian $H$ which describes the interactions between the various constituents of the system. Once $H$ is given, we can
 compute first the time evolution of the number operators as $\hat P_j(t):=e^{iHt}\hat P_j e^{-iHt}$, and then their mean values on some suitable state describing the system at $t=0$. In this way we get what we call {\em decision functions}, see formula (\ref{add1}) below. The {\em rules} needed to write down $H$ are described in \cite{bagbook}. The main idea here is that the three parties are just part of a larger system: in order to take their decisions, they need first to interact with the electors. In fact, it is mainly this interaction which creates their final decisions. Hence, $\Sc_\Pc$ must be {\em open}, i.e. there must be some environment, $\R$,  interacting with $\Pc_1$, $\Pc_2$ and $\Pc_3$, so to produce some sort of feedback used by $\Pc_j$ to decide what to do.
 The reservoir, compared with $\Sc_\Pc$, is expected to be very large, since the sets of the electors for $\Pc_1$, $\Pc_2$ and $\Pc_3$ are supposed to be rather rich. For this reason the operators of the reservoirs will be labeled also by a continuous variable\footnote{Quite often, in fact, technical reasons suggest to replace infinite sums with integrals, and this is why we use continuous variables rather than discrete indexes to label the voters.}. In other words, while the operators describing $\Pc_j$ are just three independent fermionic operators, those of the reservoirs, according to the literature on quantum open system, \cite{open}, are infinitely many fermionic operators, see (\ref{23}) and (\ref{23b})\footnote{For this reason the Hilbert space of the electors is, contrarily to $\Hil_\Pc$, infinite-dimensional.}. Once again we stress that adopting fermionic operators appears as a natural choice in our model, since it will automatically produce decision functions taking values in $[0,1]$: in this way all the relevant situations are covered, from a "will to ally", which corresponds to one, to the opposite attitude, corresponding to zero, with, of course, all the intermediate possibilities, corresponding to decision which are not {\em sharp}.

\begin{figure}
\begin{center}
\begin{picture}(450,90)

\put(160,65){\thicklines\line(1,0){45}}
\put(160,85){\thicklines\line(1,0){45}}
\put(160,65){\thicklines\line(0,1){20}}
\put(205,65){\thicklines\line(0,1){20}}
\put(183,75){\makebox(0,0){$\Pc_2$}}

\put(300,35){\thicklines\line(1,0){45}}
\put(300,55){\thicklines\line(1,0){45}}
\put(300,35){\thicklines\line(0,1){20}}
\put(345,35){\thicklines\line(0,1){20}}
\put(323,45){\makebox(0,0){$\Pc_3$}}

\put(10,35){\thicklines\line(1,0){45}}
\put(10,55){\thicklines\line(1,0){45}}
\put(10,35){\thicklines\line(0,1){20}}
\put(55,35){\thicklines\line(0,1){20}}
\put(33,45){\makebox(0,0){$\Pc_1$}}

\put(10,-55){\thicklines\line(1,0){45}}
\put(10,-35){\thicklines\line(1,0){45}}
\put(10,-55){\thicklines\line(0,1){20}}
\put(55,-55){\thicklines\line(0,1){20}}
\put(33,-45){\makebox(0,0){$\R_1$}}

\put(140,-55){\thicklines\line(1,0){85}}
\put(140,-35){\thicklines\line(1,0){85}}
\put(140,-55){\thicklines\line(0,1){20}}
\put(225,-55){\thicklines\line(0,1){20}}
\put(183,-45){\makebox(0,0){$\R_2$}}

\put(300,-55){\thicklines\line(1,0){45}}
\put(300,-35){\thicklines\line(1,0){45}}
\put(300,-55){\thicklines\line(0,1){20}}
\put(345,-55){\thicklines\line(0,1){20}}
\put(323,-45){\makebox(0,0){$\R_3$}}

\put(140,-155){\thicklines\line(1,0){85}}
\put(140,-95){\thicklines\line(1,0){85}}
\put(140,-155){\thicklines\line(0,1){60}}
\put(225,-155){\thicklines\line(0,1){60}}
\put(183,-125){\makebox(0,0){$\R_{und}$}}

\put(70,44){\thicklines\vector(1,0){220}}
\put(70,44){\thicklines\vector(-1,0){3}}
\put(70,44){\thicklines\vector(3,1){80}}
\put(70,44){\thicklines\vector(-3,-1){3}}
\put(290,44){\thicklines\vector(-3,1){80}}
\put(290,44){\thicklines\vector(3,-1){3}}

\put(31,27){\thicklines\vector(0,-1){55}}
\put(31,27){\thicklines\vector(0,1){3}}
\put(322,27){\thicklines\vector(0,-1){55}}
\put(322,27){\thicklines\vector(0,1){3}}
\put(165,57){\thicklines\vector(0,-1){85}}
\put(165,57){\thicklines\vector(0,1){3}}

\put(35,27){\thicklines\vector(1,-1){115}}
\put(35,27){\thicklines\vector(-1,1){3}}
\put(318,27){\thicklines\vector(-1,-1){115}}
\put(318,27){\thicklines\vector(1,1){3}}
\put(195,57){\thicklines\vector(0,-1){145}}
\put(195,57){\thicklines\vector(0,1){3}}


\end{picture}
 \end{center}
\vspace*{5.3cm}
\caption{\label{figscheme} The system and its multi-component reservoir.}
\end{figure}
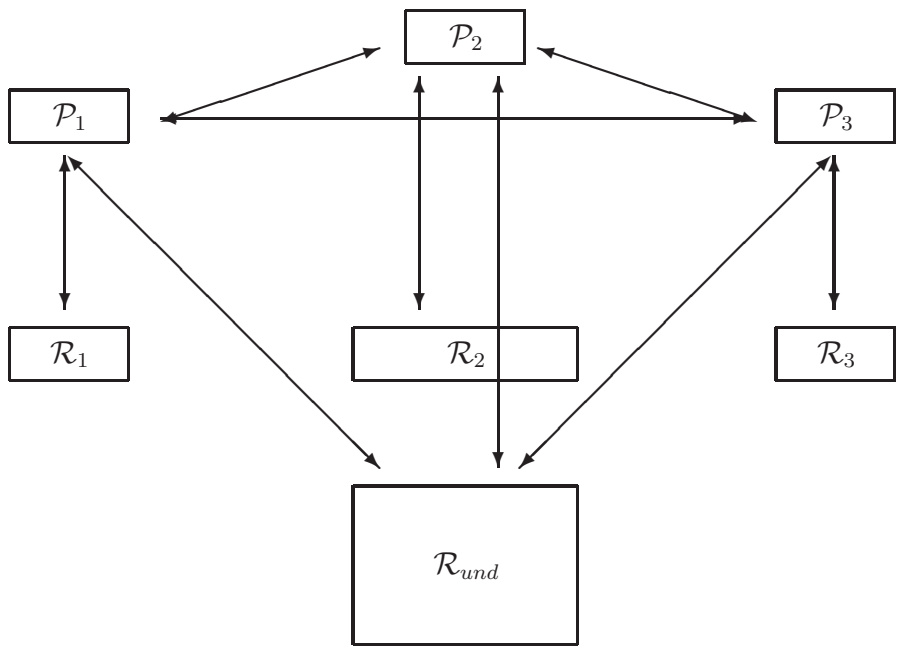

 The various elements of our model are described in Figure \ref{figscheme}, where the various arrows show all the admissible interactions.

 \vspace*{1cm}

In  Figure \ref{figscheme}, $\R_j$ represents the set of the supporters of $\Pc_j$, while $\R_{und}$ is the set of all the undecided electors. This figure also shows that, for instance, $\Pc_1$ can interact with $\R_1$ and $\R_{und}$, but not with $\R_2$ or with $\R_3$. Moreover, $\Pc_1$ interacts with both $\Pc_2$ and $\Pc_3$. The Hamiltonian which describes, in our framework, the scheme in Figure \ref{figscheme} is the following:
\be
\left\{
\begin{array}{ll}
H=H_{0}+H_{PBs}+H_{PB}+H_{int}, &  \\
H_{0}=\sum_{j=1}^{3}\omega _{j}p_j^\dagger p_j+\sum_{j=1}^{3}\int_{\Bbb R}\Omega_j(k)B_j^\dagger(k)B_j(k)\,dk+\int_{\Bbb R}\Omega(k)B^\dagger(k)B(k)\,dk,   \\
H_{PBs}=\sum_{j=1}^{3}\lambda_j\int_{\Bbb R}\left(p_j B_j^\dagger(k)+B_j(k)p_j^\dagger\right)\,dk,\\
H_{PB}=\sum_{j=1}^{3}\tilde\lambda_j\int_{\Bbb R}\left(p_j B^\dagger(k)+B(k)p_j^\dagger\right)\,dk,\\
H_{int}=\mu_{12}^{ex}\left(p_1^\dagger p_2+p_2^\dagger p_1\right)+\mu_{12}^{coop}\left(p_1^\dagger p_2^\dagger+p_2 p_1\right)+\mu_{13}^{ex}\left(p_1^\dagger p_3+p_3^\dagger p_1\right)+   \\
\qquad +\mu_{13}^{coop}\left(p_1^\dagger p_3^\dagger+p_3 p_1\right)+\mu_{23}^{ex}\left(p_2^\dagger p_3+p_3^\dagger p_2\right)+\mu_{23}^{coop}\left(p_2^\dagger p_3^\dagger+p_3 p_2\right).
\end{array}%
\right.
\label{22}\en
Here $\omega _{j}$, $\lambda_j$, $\tilde\lambda_j$, $\mu_{ij}^{ex}$ and $\mu_{ij}^{coop}$ are real quantities, while $\Omega_j(k)$ and $\Omega(k)$ are real-valued functions. We will come back on their meaning in a moment. The following CAR's for the operators of the reservoir are assumed:
\be
\{B_i(k),B_l^\dagger(q)\}=\delta_{i,l}\delta(k-q)\,\1,\qquad \{B_i(k),B_l(k)\}=0,
\label{23}
\en
as well as
\be
\{B(k),B^\dagger(q)\}=\delta(k-q)\,\1,\quad \{B(k),B(k)\}=0,
\label{23b}
\en
for all $i,l=1,2,3$, $k,\, q\in{\Bbb R}$. Moreover each $p_j^\sharp$ anti-commutes with each $B_l^\sharp(k)$ and with $B^\sharp(k)$: $\{b_j^\sharp, B_l^\sharp(k)\}=\{b_j^\sharp, B^\sharp(k)\}=0$ for all $j$, $l$ and for all $k$, and we further assume that $\{B^\sharp(q), B_l^\sharp(k)\}=0$. Here $X^\sharp$ stands for $X$ or $X^\dagger$. Assuming these CAR's is natural, since it reflects the analogous choice for the three parties.

Another reason to assume CAR's has to do with the meaning of the various terms of the Hamiltonian $H$ in (\ref{22}). Before commenting this aspect of $H$, we notice that $\lambda_j, \tilde\lambda_j$, $\mu_{jk}^{ex}$ and $\mu_{jk}^{coop}$ are all interaction parameters, measuring respectively the strength of the interaction of $\Pc_j$ with $\R_j$, with $\R_{und}$, and with the other parties $\Pc_k$. The parameters and the functions appearing in the free Hamiltonian $H_0$, $\omega_j$, $\Omega_j(k)$ and $\Omega(k)$, are related to a sort of {\em inertia} of $\Pc_j$, $\R_j$ and $\R_{und}$ respectively: in a very schematic way, we could say that the larger their values, the smaller the amplitude of the variations in time for some related dynamical variable of the model\footnote{For this reason, in analogy with classical mechanics, we have introduced the word {\em inertia} in connection with these quantities.}. This is quite a general result, suggested by many numerical and analytical computations performed along the years on similar systems, \cite{bagbook}.  $H_0$ describes the free evolution of the operators of $\Sc=\Sc_\Pc\otimes\R$, where $\R=(\R_1\otimes\R_2\otimes\R_3)\otimes\R_{und}$. If, in particular, all the interaction parameters $\lambda_j, \tilde\lambda_j$, $\mu_{ij}^{ex}$ and $\mu_{ij}^{coop}$ are zero, then $H=H_0$. Hence, since in this case $[H,\hat P_j]=0$, the number operators describing the choices of the three parties and their related decision functions stay constant in time, see Appendix. In other words, in this case the original choice of each $\Pc_j$ is not affected by the time evolution.  $H_{PBs}$, describes the interaction between the three parties and their related groups of electors, and in fact $PBs$ stands for {\em Parties $\leftrightarrow$ Backgrounds (the $\R_j$'s)} : $p_j B_j^\dagger(k)$ describes the fact that, when some sort of {\em global reaction against alliance} (GRAA)  increases, then $\Pc_j$ tends to chose 0 (no coalition). On the other hand, $B_j(k)p_j^\dagger$ describes the fact that  $\Pc_j$  tends to form a coalition when the GRAA  decreases. This is because of the raising and lowering operators $p_j^\dagger$ and $p_j$ in these interaction terms, coupled respectively with the lowering ($B_j(k)$) and raising ($B_j^\dagger(k)$) operators of the supporters of $\Pc_j$. A similar phenomenon is described by $H_{PB}$, where now $PB$ stands for {\em Parties $\leftrightarrow$ Background (i.e. $\R_{und}$)}, with the difference that the interaction is now between the parties and a single set of undecided voters. The last contribution in $H$, $H_{int}$, is introduced to describe the fact that the parties try also to talk to each other, to get some agreement. Two possibilities are allowed, one in which the parties act {\em cooperatively} (they make the same choice, and, in fact, we have terms like $p_j^\dagger p_k^\dagger$), and one in which they make opposite choices, for instance $\Pc_1$ try to form some alliance, while $\Pc_2$ excludes this possibility (and we have terms like $p_1^\dagger p_2$). Of course, the relative magnitude of $\mu_{jk}^{ex}$ and $\mu_{jk}^{coop}$ decides which is the leading contribution in $H_{int}$.

Before deducing the equations of motion for the relevant observables of the system, it is interesting to discuss the presence, or the absence, of some integrals of motion for the model. In our context, these are (self-adjoint) operators which commute with the Hamiltonian, see Appendix. In many concrete situations the existence of these kind of operators is used to suggest how the Hamiltonian should look like. Moreover, integrals of motion can also be used to check how realistic our model is, \cite{bagoliv1}. Let us introduce here
\be
\hat N =\sum_{j=1}^3 \hat N_{j}=\sum_{j=1}^3 \left(p_j^\dagger p_j+\int_{\Bbb R}B_j^\dagger(k)B_j(k)\,dk\right),
\label{25}\en
with obvious notation. It is clear that $\hat N$ is the sum of the decision operators $p_j^\dagger p_j$ plus the total GRAA.
It is easy to check that  $\left[\hat N,H\right]=0,$ if all the $\mu_{jk}^{coop}$ and the $\tilde\lambda_j$ are all zero, that is if the only allowed interaction between parties is non cooperative. On the other hand, for instance, if at least  one of the $\mu_{jk}^{coop}$'s is not zero, then $\hat N$ is no longer an integral of motion, since in this case $\left[\hat N,H\right]\neq0$: cooperation gives a non trivial dynamics to $\hat N$.

\vspace{1mm}

We can now go back to the analysis of the dynamics of the system. The Heisenberg equations of motion $\dot X(t)=i[H,X(t)]$, \cite{bagbook}, can be deduced by using the CAR's (\ref{23}) and (\ref{23b}) above:

\be
\left\{
\begin{array}{ll}
\dot p_1(t)=-i\omega_1 p_1(t)+i\lambda_1\int_{\Bbb R}B_1(q,t)\,dq+i\tilde\lambda_1\int_{\Bbb R}B(q,t)\,dq-i\mu_{12}^{ex}p_2(t)-i\mu_{12}^{coop}p_2^\dagger(t)+\\
\qquad -i\mu_{13}^{ex}p_3(t)-i\mu_{13}^{coop}p_3^\dagger(t),   \\
\vspace{1mm}
\dot p_2(t)=-i\omega_2 p_2(t)+i\lambda_2\int_{\Bbb R}B_2(q,t)\,dq+i\tilde\lambda_2\int_{\Bbb R}B(q,t)\,dq-i\mu_{12}^{ex}p_1(t)+i\mu_{12}^{coop}p_1^\dagger(t)+\\
\qquad -i\mu_{23}^{ex}p_3(t)-i\mu_{23}^{coop}p_3^\dagger(t),   \\
\vspace{1mm}
\dot p_3(t)=-i\omega_3 p_3(t)+i\lambda_3\int_{\Bbb R}B_3(q,t)\,dq+i\tilde\lambda_3\int_{\Bbb R}B(q,t)\,dq-i\mu_{13}^{ex}p_1(t)+i\mu_{13}^{coop}p_1^\dagger(t)+\\
\qquad -i\mu_{23}^{ex}p_2(t)+i\mu_{23}^{coop}p_2^\dagger(t),   \\
\vspace{1mm}
\dot B_j(q,t)=-i\Omega_j(q) B_j(q,t)+i\lambda_j p_j(t),\qquad j=1,2,3,   \\
\vspace{1mm}
\dot B(q,t)=-i\Omega(q) B(q,t)+i\sum_{j=1}^3\tilde\lambda_j p_j(t).   \label{26}
\end{array}%
\right.
\en

These last two equations can be rewritten as
$$
B_j(q,t)=B_j(q)e^{-i\Omega_j(q)t}+i\lambda_j\int_0^t p_j(t_1)e^{-i\Omega_j(q)(t-t_1)}\,dt_1
$$
and
$$
B(q,t)=B(q)e^{-i\Omega(q)t}+i\int_0^t \sum_{j=1}^3\tilde\lambda_j p_j(t_1)e^{-i\Omega(q)(t-t_1)}\,dt_1,
$$
which, assuming that $\Omega_j(k)=\Omega_j\, k$ and $\Omega(k)=\Omega\, k$, $\Omega,\Omega_j>0$,  produce
\be
\int_{\Bbb R}B_j(q,t)\,dq=\int_{\Bbb R}B_j(q)e^{-i\Omega_j q t}\,dq+i\pi\frac{\lambda_j}{\Omega_j}\,p_j(t),
\label{27}\en
and
\be
\int_{\Bbb R}B(q,t)\,dq=\int_{\Bbb R}B(q)e^{-i\Omega q t}\,dq+i\pi\frac{\sum_{j=1}^3\tilde\lambda_j\,p_j(t)}{\Omega}.
\label{28}\en
We refer to \cite{bagbook} and to \cite{open} for more details on this and similar computations. Here we just want to mention that, for instance, the choice $\Omega_j(k)=\Omega_j k$ is rather common when dealing with quantum open systems. If we now replace (\ref{27}) and (\ref{28}) in the equations (\ref{26}) for $\dot p_j(t)$, after some computations we can write them in a simple matricial form
\be
\dot q(t)=-U\,q(t)+\rho(t),
\label{29}\en
where we have defined the vectors
$$
q(t)=\left(
       \begin{array}{c}
         p_1(t) \\
         p_2(t) \\
         p_3(t) \\
         p_1^\dagger(t) \\
         p_2^\dagger(t) \\
         p_3^\dagger(t) \\
       \end{array}
     \right), \quad \rho(t)=\left(
                               \begin{array}{c}
                                 \eta_1(t) \\
                                 \eta_2(t) \\
                                 \eta_3(t) \\
                                 \eta_1^\dagger(t) \\
                                 \eta_2^\dagger(t) \\
                                 \eta_3^\dagger(t) \\
                               \end{array}
                             \right),$$ and the symmetric matrix $$U=\left(
\begin{array}{cccccc}
 \hat\omega_1 & \gamma_{1,2} & \gamma_{1,3}  & 0 & i \mu_{1,2}^{coop} & i \mu_{1,3}^{coop} \\
 \gamma_{1,2} & \hat\omega_2 & \gamma_{2,3} & -i \mu_{1,2}^{coop} & 0 & i \mu_{2,3}^{coop} \\
 \gamma_{1,3} & \gamma_{2,3} & \hat\omega_3 & -i \mu_{1,3}^{coop} & -i \mu_{2,3}^{coop} & 0 \\
 0 & -i \mu_{1,2}^{coop} & -i \mu_{1,3}^{coop} & \overline{\hat\omega_1} & \overline{\gamma_{1,2}} & \overline{\gamma_{1,3}} \\
 i \mu_{1,2}^{coop} & 0 & -i \mu_{2,3}^{coop} & \overline{\gamma_{1,2}} & \overline{\hat\omega_2} & \overline{\gamma_{2,3}} \\
 i \mu_{1,3}^{coop} & i \mu_{2,3}^{coop} & 0 & \overline{\gamma_{1,3}} & \overline{\gamma_{2,3}} & \overline{\hat\omega_3} \\
\end{array}
\right).
$$
Here we have introduced the following simplifying notation:
$$
\mu_l:=\frac{\lambda_l^2}{\Omega_l}+\frac{\tilde\lambda_l^2}{\Omega}, \quad \hat\omega_l:=i\omega_l+\pi\mu_l, \quad  \gamma_{k,l}:=i\mu_{k,l}^{ex}+\frac{\pi}{\Omega}\tilde\lambda_k\tilde\lambda_l,
$$
for $k,l=1,2,3$, as well as the operator-valued functions:
$$
\eta_j(t)=i\left(\lambda_j\beta_j(t)+\tilde\lambda_j\beta(t)\right), \quad\beta_j(t)=\int_{\Bbb R}B_j(q)e^{-i\Omega_j q t}dq, \quad \beta(t)=\int_{\Bbb R}B(q)e^{-i\Omega q t}dq.
$$
The solution of (\ref{29}) is easily found in a matricial form:
\be
q(t)=e^{-U\,t}q(0)+\int_0^t e^{-U\,(t-t_1)}\,\rho(t_1)\,dt_1,
\label{210}\en
which is now the starting point for our analysis below.

\vspace{2mm}

{\bf Remark:--} a particularly simple situation occurs when there is no interaction at all, i.e. when $\lambda_j=\tilde\lambda_j=\mu_{k,l}^{coop}=\mu_{k,l}^{ex}=0$, for all $j,k,l=1,2,3$. In this case, we trivially have $H=H_0$ and no interesting dynamics is expected. Indeed, this is reflected by the fact that $U$ becomes a diagonal matrix with purely imaginary elements, and the equations for each fermionic mode produce oscillations for the creation and annihilation operators $p_j^\dagger(t)$ and $p_j(t)$, and constant values for their products, the number operators $\hat P_j(t)=p_j^\dagger(t)p_j(t)$, $j=1,2,3$. Because of our definition (\ref{add1}) below, in this case all the decision functions are constant in time. This is reasonable, since $\Pc_1$, $\Pc_2$ and $\Pc_3$ feel no interaction at all.

\vspace{2mm}

Once we have obtained $q(t)$, we need to compute the decision functions $P_j(t)$, which are defined as follows:  \be P_j(t):=\left<\hat P_j(t)\right>=\left<p_j^\dagger(t)p_j(t)\right>,\label{add1}\en $j=1,2,3$. Here $\left<.\right>$ is a state over the full system. These states, \cite{bagbook}, are taken to be suitable tensor products of vector states on $\Sc_\Pc$ and states on the reservoir which obey some standard rules, see below. More in details,  for each operator of the form $X_{\Sc}\otimes Y_{\R}$, $X_{\Sc}$ being an operator of $\Sc_\Pc$ and $Y_{\R}$ an operator of
the reservoir, we put
\be
\left\langle X_{\Sc}\otimes Y_{\R}\right\rangle :=\left\langle \varphi_{n_1,n_2,n_3},X_{\Sc}\varphi_{n_1,n_2,n_3}\right\rangle \,\omega
_{\R}(Y_{\R}).
\label{add2}\en
Here $\varphi_{n_1,n_2,n_3}$ is one of the vectors introduced at the beginning of this section, and each $n_j$ represents, as discussed before, the tendency of $\Pc_j$ to form or not some coalition at $t=0$. Moreover, $\omega _{\R}(.)$ is a state on $\R$ satisfying
the following standard properties, \cite{bagbook}:
\be
\omega _{\R}(1\!\!1_{\R})=1,\quad \omega _{\R}(B_{j}(k))=\omega
_{\R}(B_{j}^{\dagger }(k))=0,\quad \omega _{\R}(B_{j}^{\dagger
}(k)B_{l}(q))=N_{j}(k)\,\delta _{j,l}\delta (k-q),
\label{211}\en
as well as
\be
\omega _{\R}(B(k))=\omega
_{\R}(B^{\dagger }(k))=0,\quad \omega _{\R}(B^{\dagger
}(k)B(q))=N(k)\,\delta (k-q),
\label{211bis}\en
for some suitable functions $N_{j}(k)$, $N(k)$ which we take here to be constant in $k$: $N_{j}(k)=N_j$ and $N(k)=N$.  Also, we assume that $\omega
_{\R}(B_{j}(k)B_{l}(q))=\omega
_{\R}(B(k)B(q))=0$, for all $j$ and $l$. In our framework, the state in (\ref{add2}) describes the fact that, at $t=0$,  $\Pc_j$'s decision (concerning alliances) is $n_j$, while the overall feeling of the voters $\R_j$ is $N_j$, and that of the undecided ones is $N$. Of course, these might appear as oversimplifying assumptions, and in fact they are, but still they produce, in many concrete applications, a rather interesting dynamics for the model.

Let us now call $V_t:=e^{-Ut}$, and $(V_t)_{j,k}$ its $(j,k)$-th matrix element. Then some long but straightforward computations produce the following result:
\be
P_j(t)=P_j^{(a)}(t)+P_j^{(b)}(t),
\label{212}\en
with
$$
P_j^{(a)}(t)=\sum_{k=1}^3\left(\left|(V_t)_{j,k}\right|^2n_k+\left|(V_t)_{j,k+3}\right|^2(1-n_k)\right)
$$
and
$$
P_j^{(b)}(t)= 2\pi\int_0^tdt_1\sum_{k=1}^3\left(p_k^{(j)}(t-t_1)M_k+p_{k+3}^{(j)}(t-t_1)M_k^c\right) +
$$
$$
+2\pi\int_0^tdt_1\sum_{k, l=1,\, k<l}^3\left(p_{k,l}^{(j)}(t-t_1)\theta_{k,l}+p_{3+k,3+l}^{(j)}(t-t_1)\theta_{k,l}^c\right),
$$
$j=1,2,3$, where we have also introduced the shorthand notation $M_j:=\frac{\lambda_j^2N_j}{\Omega_j}+\frac{\tilde\lambda_j^2N}{\Omega}$, $M_j^c:=\frac{\lambda_j^2(1-N_j)}{\Omega_j}+\frac{\tilde\lambda_j^2(1-N)}{\Omega}$ as well as $\theta_{k,l}=\tilde\lambda_k\tilde\lambda_l\,\frac{N}{\Omega}$ and $\theta_{k,l}^c=\tilde\lambda_k\tilde\lambda_l\,\frac{1-N}{\Omega}$, for $j=1,2,3$ and $k,l=1,2,3$ with $k<l$. We have also defined the following functions:
$$
p_k^{(j)}(t)=\left|(V_t)_{j,k}\right|^2, \qquad p_{k,l}^{(j)}(t)=2\Re\left[\overline{(V_t)_{j,k}}\,(V_t)_{j,l}\right],
$$
where $\Re(z)$ stands for the real part of the complex quantity $z$.
\vspace{2mm}

{\bf Remark:--} In formula (\ref{212}) we have divided the dependence of the decision functions $P_j(t)$ in two parts: $P_j^{(a)}(t)$ contains all the contributions coming from $\Sc_\Pc$, while $P_j^{(b)}(t)$ contains the contributions coming from the electors. We see that these two kind of contributions differ essentially for the presence of some integrations in $P_j^{(b)}(t)$, while no time integral appears in $P_j^{(a)}(t)$. This has interesting consequences when computing the asymptotic values of the decision functions, as, for instance,  formulas (\ref{32}) and (\ref{33}) below clearly show. We will see that, in both those formulas,  $P_j^{(a)}(t)\rightarrow0$ when $t\rightarrow\infty$, while $P_j^{(b)}(t)$ does not tend to zero. In fact, this is rather general: the only non trivial contributions in $P_j(t)$, for large $t$, comes always from the integrals in $P_j^{(b)}(t)$ and not from $P_j^{(a)}(t)$. In other words, the final decision is always a consequence of {\em the whole story}. This will be made more evident in Sections III and IV.

\section{The parties do not talk to each other}

We start considering a (sadly) realistic situation, i.e. the case in which the parties do not talk to each other, and they only talk to their own supporters. In other words, each $\Pc_j$ only interacts with $\R_j$, but not with $\R_{und}$ or among them. Later in this section we will see what happens when one, two and all the parties interact also with $\R_{und}$ while they still do not talk to each other. Except for this last case, a reasonably simple expression for the functions $P_j(t)$ can be deduced analytically. However, sometimes it is convenient (see Section \ref{sect2partiti}) to use a perturbative expansion in the interaction parameters\footnote{which, by the way, only works up to a certain extent.}, while in more general situations considered in this paper, see Section \ref{sect3partiti} and Section \ref{sect4}, it is surely more convenient to use numerical techniques.

\subsection{Case 1: {\em almost} no interaction}

Since the parties do not talk to each other, we have to put $\mu_{k,l}^{ex}=\mu_{k,l}^{coop}=0$ in (\ref{22}) for all $k$ and $l$. Furthermore, since they also don't interact with $\R_{und}$, we have to fix $\tilde\lambda_k=0$, $k=1,2,3$. In this case, the matrix $U$ is particularly simple since it becomes diagonal, with $\hat\omega_l=i\omega_l+\pi\frac{\lambda_l^2}{\Omega_l}$. Then $V_t$ is also diagonal, with obvious matrix elements. As a result, formula (\ref{212}) simplifies significantly:
\be
P_j(t)=\left|(V_t)_{j,j}\right|^2n_j+2\pi\,M_j\,\int_0^tdt_1\,\left|(V_{t-t_1})_{j,j}\right|^2,
\label{31}\en
$j=1,2,3$. We see that, not surprisingly, all the parties behave in a very similar way: this is natural, because there is no difference between $\Pc_1$, $\Pc_2$ and $\Pc_3$, except, at most, in the numerical values of their related parameters, and because each subsystem $(\Pc_j,\R_j)$ is independent of the others. Computing the integral in (\ref{31}), with simple algebraic manipulations we deduce
\be
P_j(t)=n_j\,e^{-2\pi\,t\,\frac{\lambda_j^2}{\Omega_j}}+N_j\left(1-e^{-2\pi\,t\,\frac{\lambda_j^2}{\Omega_j}}\right),
\label{32}\en
which goes to $N_j$ when $t$ diverges: $P_j(\infty):=\lim_{t\rightarrow\infty}P_j(t)=N_j$, $j=1,2,3$. The conclusion is simple: in this case each party does what its electors decide. In other words: $\Pc_1$ just does not care about other opinions, except those of $\R_1$. This, again, appears quite reasonable. It is also interesting to notice that the speed of convergence of $P_j(t)$ to its asymptotic value depends on the ratio $\frac{\lambda_j^2}{\Omega_j}$: the higher this ratio, the higher this speed. This means that the strength of the interaction $\Pc_j\leftrightarrow\R_j$, measured by $\lambda_j$, is relevant also to determine the {\em speed of decision}.

\subsection{Case 2:  $\Pc_1$ interacts with $\R_{und}$}

This is again a simple situation, from an analytical point of view. In fact, since the parties do not talk to each other, we have again $\mu_{k,l}^{ex}=\mu_{k,l}^{coop}=0$ for all $k$ and $l$. On the other hand,  $\tilde\lambda_1\neq0$, but we still have $\tilde\lambda_2=\tilde\lambda_3=0$. This implies that, because of the definition of $\gamma_{k,l}$, $U$ is again a diagonal matrix. As in the previous case we have
$$
U=diag\left(\hat\omega_1,\hat\omega_2,\hat\omega_3,\overline{\hat\omega_1},\overline{\hat\omega_2},\overline{\hat\omega_3}\right),
$$
and $V_t$ is a diagonal matrix as well. The difference with the previous case arises because of the fact that $\tilde\lambda_1\neq0$. Repeating the same steps as above, we deduce the following analytic expression for $P_j(t)$:
\be
P_j(t)=n_j\,e^{-2\pi\,t\,\mu_j}+\frac{M_j}{\mu_j}\,\left(1-e^{-2\pi\,t\mu_j}\right),
\label{33}\en
$j=1,2,3$, which implies, first of all, that  $P_2(\infty)=N_2$ and $P_3(\infty)=N_3$. This is expected since there is no difference concerning $\Pc_2$ and $\Pc_3$ with respect to what happens in Section III.1. Moreover, we also deduce that $$P_1(\infty)=\frac{M_1}{\mu_1}=\frac{\frac{\lambda_1^2 N_1}{\Omega_1}+\frac{\tilde\lambda_1^2 N}{\Omega}}{\frac{\lambda_1^2 }{\Omega_1}+\frac{\tilde\lambda_1^2 }{\Omega}},$$
which means that, this time, the final decision of $\Pc_1$ is influenced also by by $\R_{und}$, i.e. by $\tilde\lambda_1$, $N$ and $\Omega$. This dependence is somehow masked if $\R_1$ and $\R_{und}$ {\em share the same opinion}, i.e. if $N_1=N$. In this case, in fact,  $P_1(\infty)=N_1$, as if there were no $\R_{und}$ at all. This can be easily understood since, if $N_1=N$, then, even if $\R_1$ and $\R_{und}$ are groups of different people, still they all have the same opinion on what $\Pc_1$ should do. On the other hand, if $N_1\neq N$, it is easy to see that $P_1(\infty)\in ]0,1[$: the final decision is  not sharp now, contrarily to what happens when $\tilde\lambda_1=0$ (when $P_1(\infty)=N_1$), and it depends on the ratio between $\tilde\lambda_1^2\Omega_1$ and $\lambda_1^2\Omega$. In fact, fixing  for instance $N_1=1$ and $N=0$, we get
\be
P_1(\infty)=\frac{1}{1+\frac{\tilde\lambda_1^2\Omega_1}{\lambda_1^2\Omega}},
\label{31bis}\en
which shows that the presence of $\R_{und}$ modifies the otherwise clear attitude of $\Pc_1$. In fact, since $N_1=1$ here, $\Pc_1$ would try to ally with some other party. However, since $N=0$, the electors in $\R_{und}$ would prefer that $\Pc_1$ does not form any coalition, and, as a consequence, the result is a number which is neither zero nor one. However, if for instance $\tilde\lambda_1^2\Omega_1\ll\lambda_1^2\Omega$, then $P_1(\infty)$ approaches one. This is what we expect since, in this case, at least if $\Omega$ and $\Omega_1$ are of the same order of magnitude, the interaction between $\Pc_1$ and $\R_1$ is much stronger than the one between $\Pc_1$ and $\R_{und}$.  Similar conclusions can be deduced if $N_1=0$ and $N=1$.

\vspace{2mm}

{\bf Remark:--} Similarly, if we assume that $\Pc_2$ (rather than $\Pc_1$) is the only party talking with $\R_{und}$, we would get a similar result: $P_1(\infty)=N_1$, $P_3(\infty)=N_3$, while $P_2(\infty)=\frac{M_2}{\mu_2}$.

\subsection{Case 3: $\Pc_1$ and $\Pc_2$ interact weakly with $\R_{und}$}\label{sect2partiti}

Once again we assume that $\mu_{k,l}^{ex}=\mu_{k,l}^{coop}=0$ for every $k$ and $l$. On the other hand, we put now  $\tilde\lambda_1\neq0$, $\tilde\lambda_2\neq0$, and $\tilde\lambda_3=0$. Then $U$ is no longer a diagonal matrix. Indeed it looks like
$$U=\left(
\begin{array}{cccccc}
 \hat\omega_1 & \gamma_{1,2} & 0  & 0 & 0 & 0 \\
 \gamma_{1,2} & \hat\omega_2 & 0 & 0 & 0 & 0 \\
 0 & 0 & \hat\omega_3 & 0 & 0 & 0 \\
 0 & 0 & 0 & \overline{\hat\omega_1} & {\gamma_{1,2}} & 0 \\
 0 & 0 & 0 & {\gamma_{1,2}} & \overline{\hat\omega_2} & 0 \\
 0 & 0 & 0 & 0 & 0 & \overline{\hat\omega_3} \\
\end{array}
\right),
$$

\vspace*{2mm}

\noindent and $V_t$ has a similar expression: the only non diagonal elements of $V_t$ which are non zero are those with entries 12, 21, 45 and 54. Notice also that $V_t$ is symmetric. In view of the fact that, in this section, we want to deduce reasonably simple analytical results, we start working under the assumption that $\max\{\tilde\lambda_1,\tilde\lambda_2\}\ll\min\{\lambda_1,\lambda_2,\lambda_3\}$: hence we have interactions with $\R_{und}$, but these are weak with respect to those $\Pc_1$, $\Pc_2$ and $\Pc_3$ have with their own electors. So {\em electors come first!} To simplify the notation, we also fix here $\Omega_1=\Omega_2=\Omega=1$. Under these conditions it is possible to deduce an analytic expression for the matrix elements of $V_t$. In particular, the diagonal terms are always the same: $(V_t)_{jj}=e^{-\hat\omega_j t}$ if $j=1,2,3$ and $(V_t)_{jj}=e^{-\overline{\hat\omega_{j-3}} t}$ if $j=4,5,6$. The other non zero terms depend on whether $\hat\omega_1$ is equal to $\hat\omega_2$ or not. We restrict here to this first case, since explicit formulas are simpler. In this case we have, for instance
$$
P_1^{(a)}(t)=\left|(V_t)_{1,1}\right|^2n_1+\left|(V_t)_{1,2}\right|^2n_2,
$$
where
$$
\left|(V_t)_{1,1}\right|^2=\left|e^{-\hat\omega_1 t}\right|^2=e^{-2\pi(\lambda_1^2+\tilde\lambda_1^2)t}, \qquad \left|(V_t)_{1,2}\right|^2=\gamma_{1,2}^2 t^2\left|e^{-\hat\omega_1 t}\right|^2=\gamma_{1,2}^2 t^2 e^{-2\pi(\lambda_1^2+\tilde\lambda_1^2)t}.
$$
Then
$$
P_1^{(a)}(t)=e^{-2\pi(\lambda_1^2+\tilde\lambda_1^2)t}\left(n_1+n_2 \gamma_{1,2}^2 t^2\right)\rightarrow 0,
$$
for $t\rightarrow\infty$. Hence we deduce also here that the contribution of $P_1^{(a)}(t)$ to the decision function for $\Pc_1$ disappears for $t$ very large. Slightly more complicated is the computation of $P_1^{(b)}(\infty)=\lim_{t,\infty}P_1^{(b)}(t)$. With obvious notation we get:
$$
P_1(\infty)=P_1^{(a)}(\infty)+P_1^{(b)}(\infty)\simeq N_1+\left(\frac{\tilde\lambda_1}{\lambda_1}\right)^2N+\frac{(\tilde\lambda_1\tilde\lambda_2)^2}{\lambda_2^4}\left[\frac{\lambda_2^2N_2+\tilde\lambda_2^2N}{2\lambda_2^2}-N\right].
$$

\vspace{2mm}

A similar result can be deduced for $P_2(\infty)$. Indeed we get:
$$
P_2(\infty)=P_2^{(a)}(\infty)+P_2^{(b)}(\infty)\simeq N_2+\left(\frac{\tilde\lambda_2}{\lambda_2}\right)^2N+\frac{(\tilde\lambda_1\tilde\lambda_2)^2}{\lambda_1^4}\left[\frac{\lambda_1^2N_1+\tilde\lambda_1^2N}{2\lambda_1^2}-N\right].
$$
On the other hand, due to the fact that we have fixed $\tilde\lambda_3=0$, nothing changes for the asymptotic value of $P_3(t)$: $P_3(\infty)=N_3$.

The above formulas for $P_1(\infty)$ and $P_2(\infty)$ show a strange feature, which is due to the perturbative expansion considered here: suppose $N=0$. Then $P_1(\infty)\simeq N_1+\frac{(\tilde\lambda_1\tilde\lambda_2)^2}{2\lambda_2^4}\,N_2$, and $P_2(\infty)\simeq N_2+\frac{(\tilde\lambda_1\tilde\lambda_2)^2}{2\lambda_1^4}\,N_1$. Therefore, because of our approximation scheme, it may happen that $P_j(\infty)$ is slightly larger than one. This means that these perturbative results must be taken {\em cum grano salis}, since we must always have $P_j(t)\in[0,1]$ for all $j$ and for all $t$. Hence, the conclusion is that the perturbative expansion proposed here can only give some suggestions of what is going on, but not rigorous results. Nevertheless, at least in principle, the exact solution could still be found, due to the fact that the model is linear, but we will not give its very complicated analytic expression here, since numerical results are sufficient for us. We just recall that this analytical solution can be deduced from (\ref{210}).

\begin{figure}[ht]
\begin{center}
\includegraphics[width=0.4\textwidth]{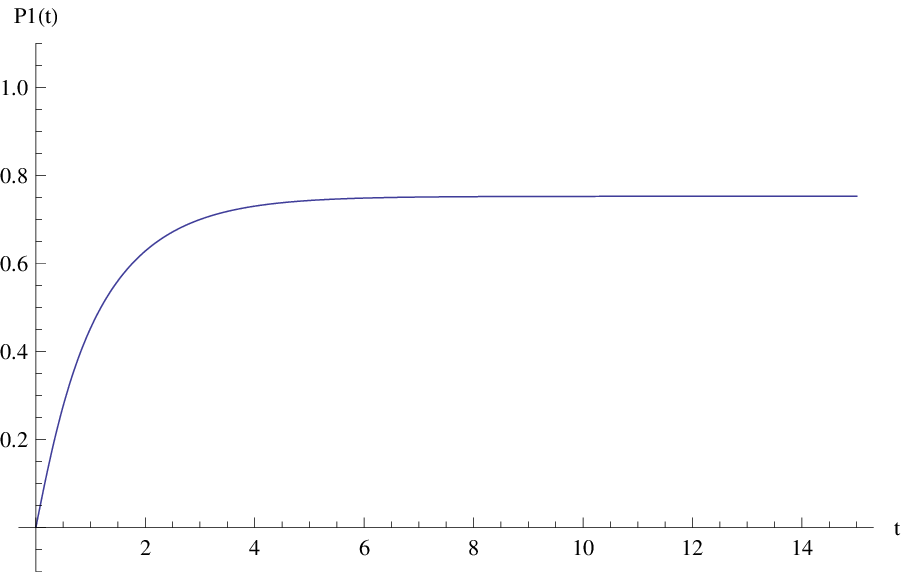}
\includegraphics[width=0.4\textwidth]{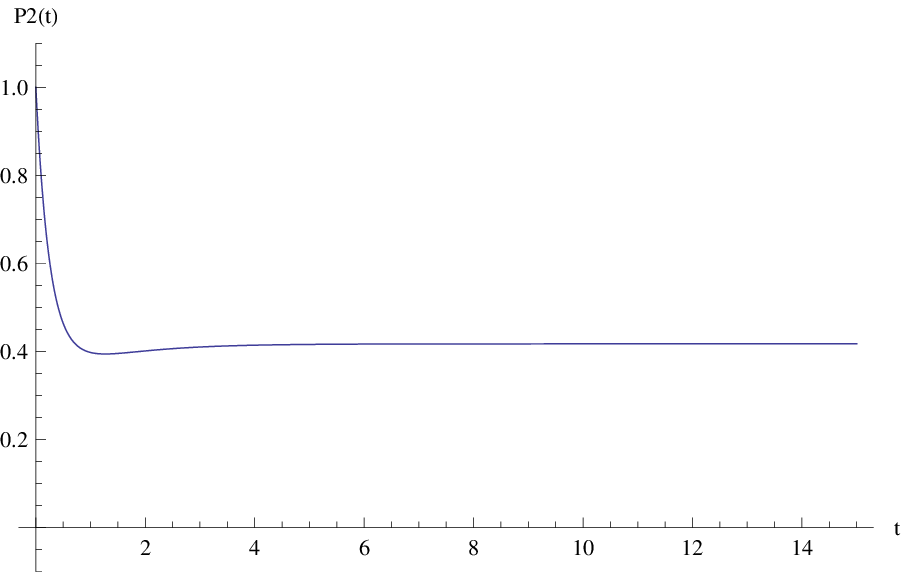}\hfill\\[0pt]
\includegraphics[width=0.4\textwidth]{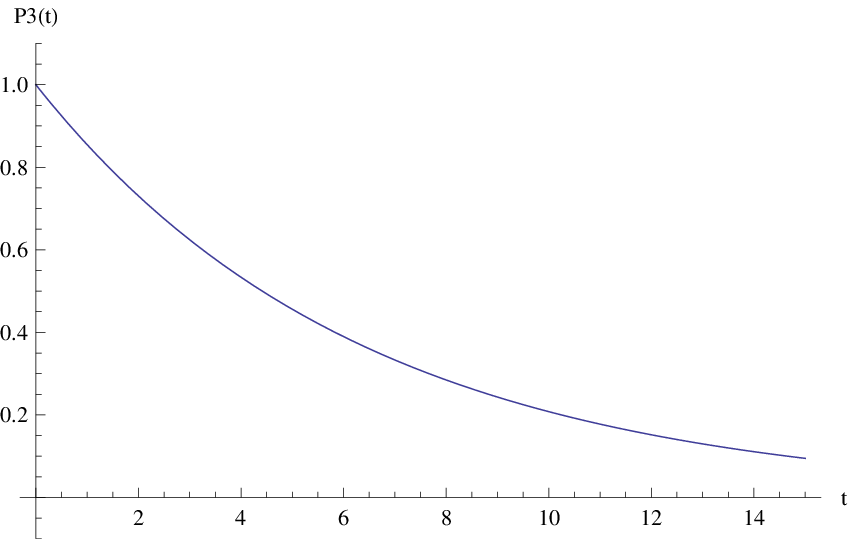}
\end{center}
\caption{{\protect\footnotesize $P_1(t)$ (top left), $P_2(t)$ (top right) and $P_3(t)$ (bottom) for $\mu_{k,l}^{ex}=\mu_{k,l}^{coop}=0$, $\omega_1=1$, $\omega_2=\omega_3=2$, $
\Omega_1=\Omega_3=\Omega=0.1$, $\Omega_2=0.2$, $\lambda_1=0.1$, $\lambda_2=0.2$, $\lambda_3=0.05$, $\tilde\lambda_1=0.1$,
$\tilde\lambda_2=0.2$, $\tilde\lambda_3=0$, and $n_1=0$, $n_2=n_3=1$, $N_1=N_2=1$, $N_3=N=0$.}}
\label{fig1}
\end{figure}

A bit more complicated is to deduce an analytical result when $\hat\omega_1\neq\hat\omega_2$, while the numerical plots can again be easily drawn  in this case.
For instance, in Figure \ref{fig1} we plot the decision functions $P_j(t)$, $j=1,2,3$, for a particular choice of the parameters of the Hamiltonian and for a particular choice of the initial conditions, i.e. of $n_j$, $N_j$ and $N$. Similar plots can be obtained for different choices of the parameters and of the initial conditions.

 From Figure \ref{fig1} we see that $\Pc_1$ and $\Pc_2$, because of the interactions with $\R_j$ and $\R_{und}$, modify their original attitudes, going to a sort of intermediate state (i.e., they are not able to get a sharp decision). After some short transient, $\Pc_1$ appears quite interested in forming some coalition, while $\Pc_2$ loses (part of) its original interest very soon. The situation is even more extreme for $\Pc_3$, which does not interact with $\R_{und}$ but only with $\R_3$. In this case, $\Pc_3$ completely modifies his original attitude, following the mood of its supporters, $\R_3$ (Notice in fact that, in Figure 2, $N_3=0$). We see from the plots that the interaction with $\R_{und}$ can produce some uncertainty  in the final decision.


\subsection{Case 4: all the parties interact (also) with $\R_{und}$}\label{sect3partiti}
\begin{figure}[ht]
\begin{center}
\includegraphics[width=0.4\textwidth]{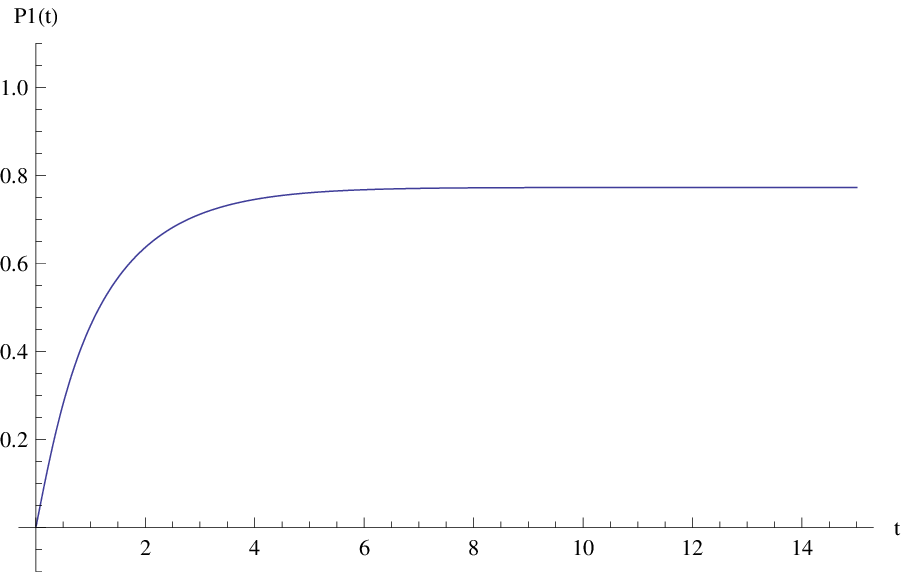}\hspace{8mm} %
\includegraphics[width=0.4\textwidth]{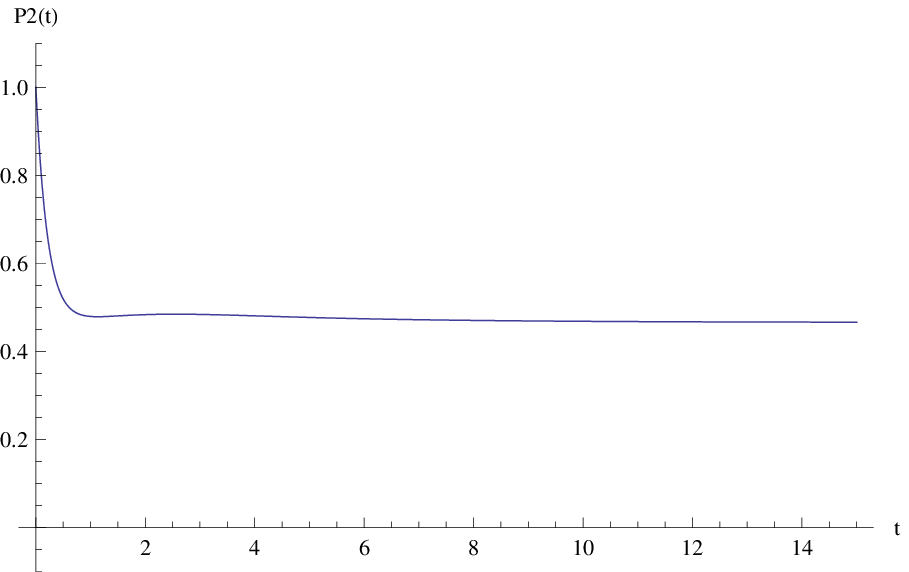}\hfill\\[0pt]
\includegraphics[width=0.4\textwidth]{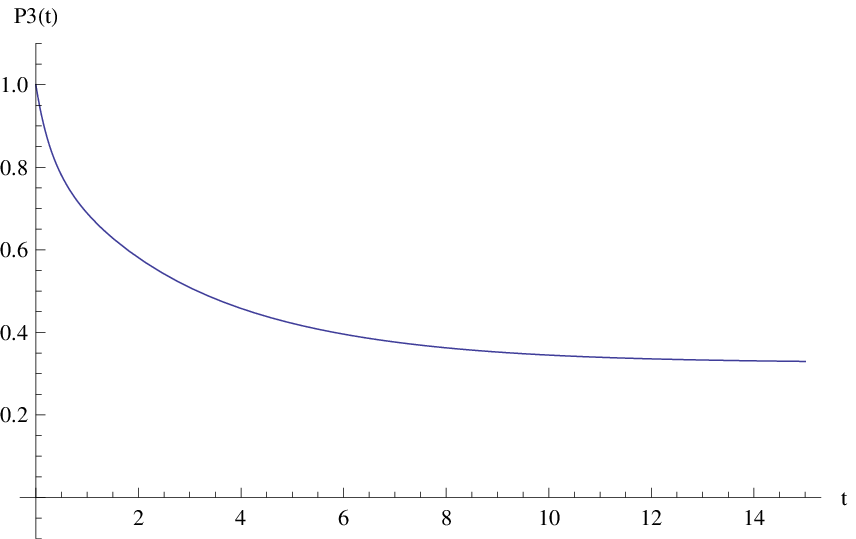}
\end{center}
\caption{{\protect\footnotesize $P_1(t)$ (top left), $P_2(t)$ (top right) and $P_3(t)$ (bottom) for $\mu_{k,l}^{ex}=\mu_{k,l}^{coop}=0$, $\omega_1=1$, $\omega_2=\omega_3=2$, $
\Omega_1=\Omega_3=\Omega=0.1$, $\Omega_2=.2$, $\lambda_1=0.1$, $\lambda_2=0.2$, $\lambda_3=0.05$, $\tilde\lambda_1=0.1$,
$\tilde\lambda_2=0.2$, $\tilde\lambda_3=0.1$, and $n_1=0$, $n_2=n_3=1$, $N_1=N_2=1$, $N_3=N=0$.}}
\label{fig2}
\end{figure}

We conclude this section considering what happens if, again, $\mu_{k,l}^{ex}=\mu_{k,l}^{coop}=0$ for all $k$ and $l$, but  $\tilde\lambda_j\neq0$, $j=1,2,3$. The parties do not talk to each other, but they all communicate with their own electors and with the undecided voters. In this case $U$ looks like
$$U=\left(
\begin{array}{cccccc}
 \hat\omega_1 & \gamma_{1,2} & \gamma_{1,3}  & 0 & 0 & 0 \\
 \gamma_{1,2} & \hat\omega_2 & \gamma_{2,3} & 0 & 0 & 0 \\
 \gamma_{1,3} & \gamma_{2,3} & \hat\omega_3 & 0 & 0 & 0 \\
 0 & 0 & 0 & \overline{\hat\omega_1} & {\gamma_{1,2}} & \gamma_{1,3} \\
 0 & 0 & 0 & {\gamma_{1,2}} & \overline{\hat\omega_2} & \gamma_{2,3} \\
 0 & 0 & 0 & \gamma_{1,3} & \gamma_{2,3} & \overline{\hat\omega_3} \\
\end{array}
\right).
$$
In Figure \ref{fig2} we plot the decision functions $P_j(t)$ using essentially the same values of the parameters as in Figure \ref{fig1}, except for $\tilde\lambda_3$, which here is taken positive ($\tilde\lambda_3=0.1$). As we see, the only major difference between Figures \ref{fig1} and \ref{fig2} is in the third function, $P_3(t)$, which does not decay to zero, but goes to a positive asymptotic value: because of the interaction with $\R_{und}$, there is some chance for a coalition, now.

Considering other values of the initial conditions, and in particular of $N_3$, we arrive to the following rather general conclusion: when $\tilde\lambda_3=0$, then $P_3(\infty)$ approaches, in all the cases considered in our analysis, the value $N_3$, so it can only be zero or one. On the other hand, if $\tilde\lambda_3\neq0$, then $P_3(\infty)$ approaches $N_3$, but {\em not so much}, and in fact the numerical values we obtain are always between zero and one. The decision of $\Pc_3$ is driven by $\R_3$, but not only: the interaction with $\R_{und}$ makes the decision slightly more flexible. This is the same effect we have already observed, for instance,  in Section III.2, and, in fact, can be seen as a general result.

\section{What happens when the parties talk to each other?}\label{sect4}

Our next step is to consider what happens to the decision functions $P_j(t)$ when we also allow the parties to interact among them. This might occur far before the election, in that temporal window in which, apparently, all the politicians seem to be interested in finding some agreement with their competitors. This window, usually, does not last long: when the election day approaches, then usually  each party tends to attack the other parties more and more. Moreover, after the elections, the winner might be interested in collaborating with other parties only if it has not the majority in the Parliament. Otherwise, usually the party who won the elections  is not interested in any alliance at all.  From a mathematical point of view, the main difference here with respect to what we have done in Section III is that we will now assume that $\mu_{k,l}^{ex}$, or $\mu_{k,l}^{coop}$, or both, are non zero.

\subsection{No cooperative effect}

\begin{figure}[ht]
\begin{center}
\includegraphics[width=0.4\textwidth]{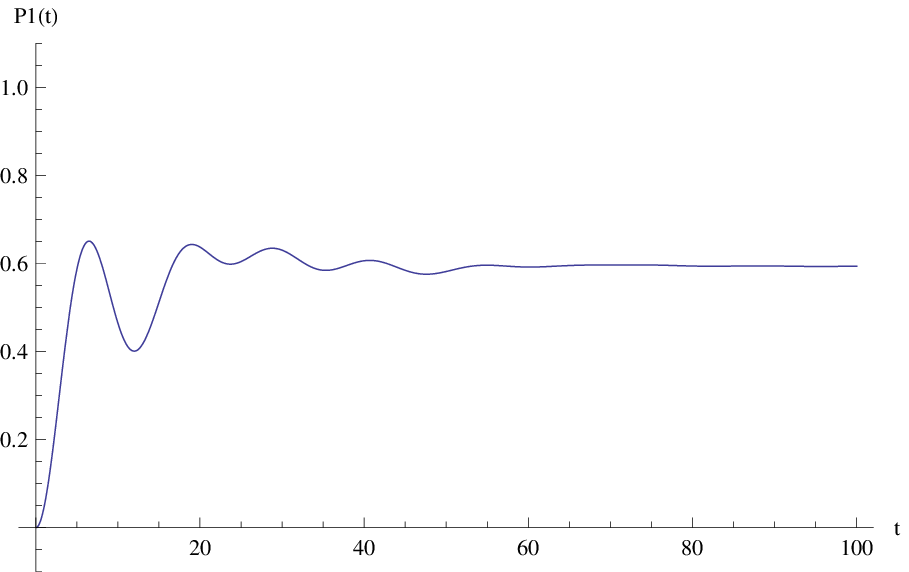}\hspace{8mm} %
\includegraphics[width=0.4\textwidth]{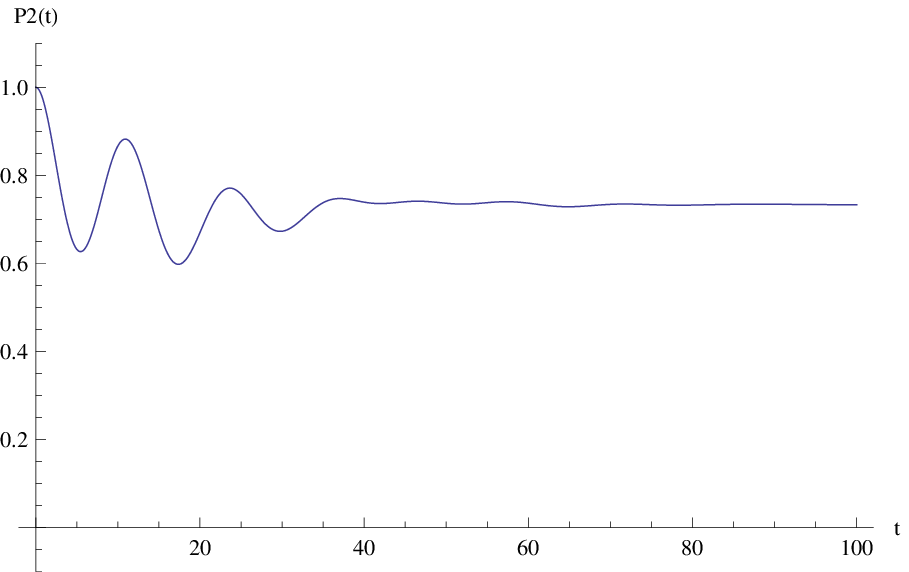}\hfill\\[0pt]
\includegraphics[width=0.4\textwidth]{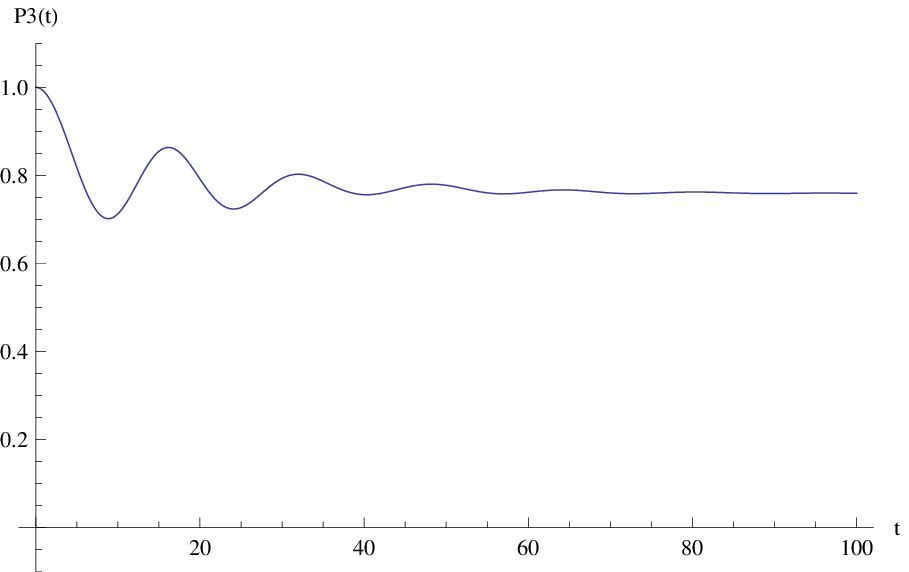}
\end{center}
\caption{{\protect\footnotesize $P_1(t)$ (top left), $P_2(t)$ (top right) and $P_3(t)$ (bottom) for $\mu_{1,2}^{ex}=0.2$, $\mu_{1,3}^{ex}=0.1$, $\mu_{2,3}^{ex}=0.15$, $\mu_{k,l}^{coop}=\tilde\lambda_j=0$, $\omega_1=0.1$, $\omega_2=\omega_3=0.2$, $
\Omega_1=\Omega_3=1$, $\Omega_2=2$, $\Omega=0.1$, $\lambda_1=0.1$, $\lambda_2=0.2$, $\lambda_3=0.05$, and $n_1=0$, $n_2=n_3=1$, $N_1=0$, $N_2=N_3=N=1$.}}
\label{fig3}
\end{figure}

\begin{figure}[ht]
\begin{center}
\includegraphics[width=0.4\textwidth]{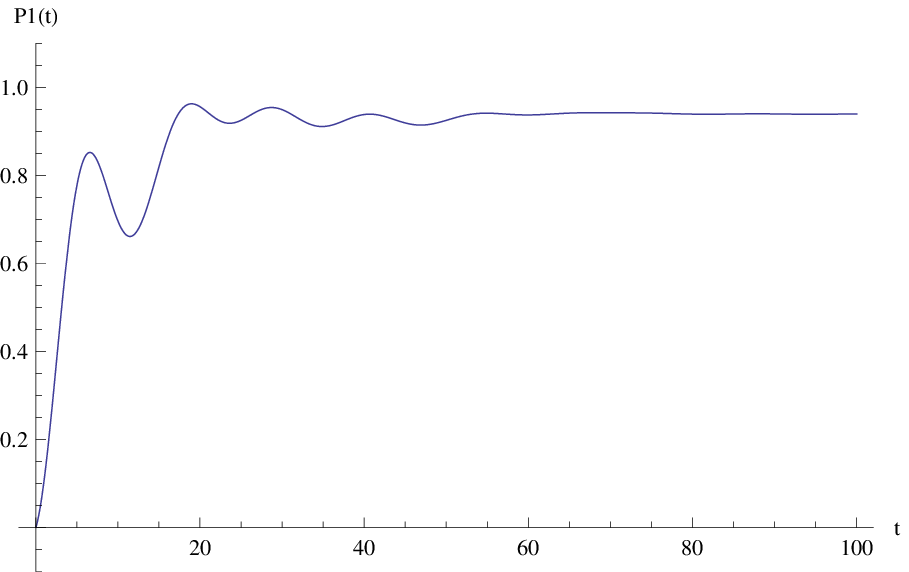}\hspace{8mm} %
\includegraphics[width=0.4\textwidth]{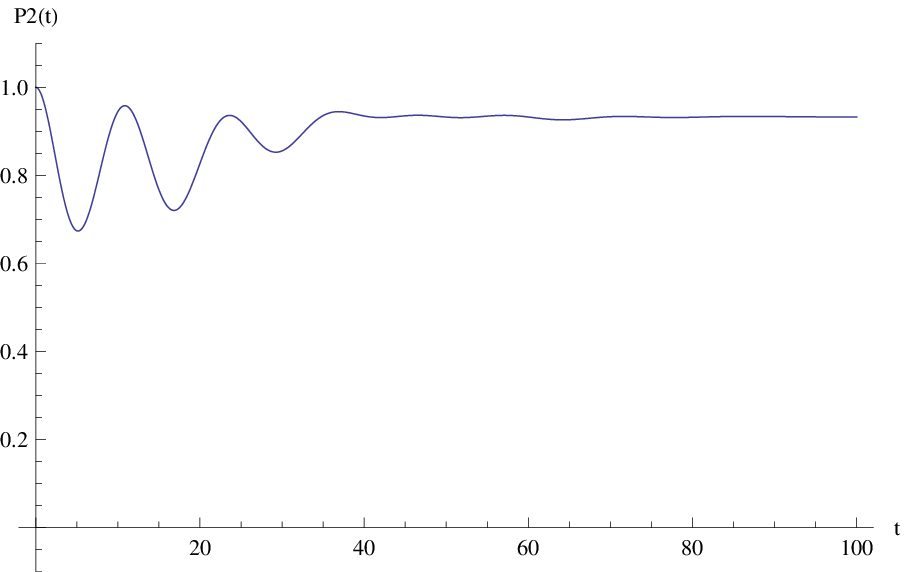}\hfill\\[0pt]
\includegraphics[width=0.4\textwidth]{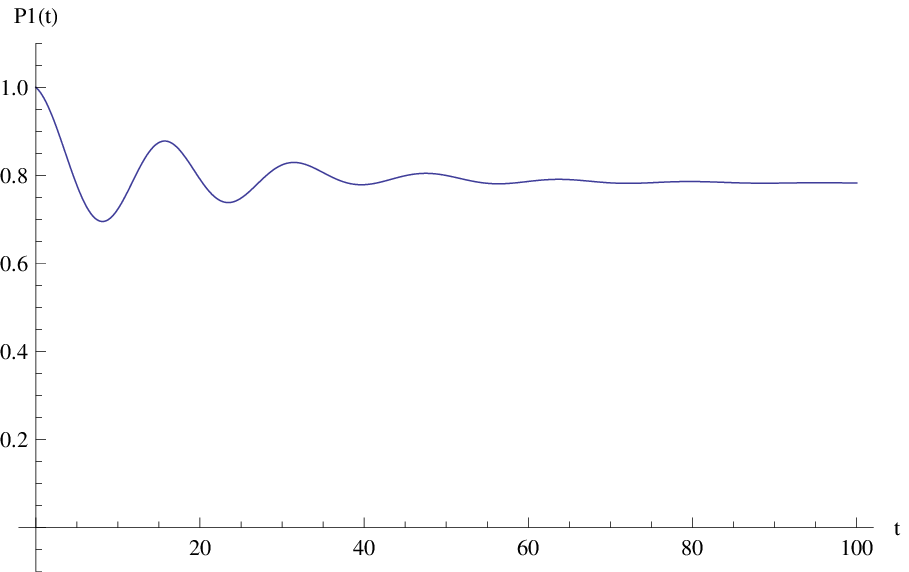}
\end{center}
\caption{{\protect\footnotesize $P_1(t)$ (top left), $P_2(t)$ (top right) and $P_3(t)$ (bottom) for $\mu_{1,2}^{ex}=0.2$, $\mu_{1,3}^{ex}=0.1$, $\mu_{2,3}^{ex}=0.15$, $\mu_{k,l}^{coop}=\tilde\lambda_j=0$, $\omega_1=0.1$, $\omega_2=\omega_3=0.2$, $
\Omega_1=\Omega_3=1$, $\Omega_2=2$, $\Omega=0.1$, $\lambda_1=0.1$, $\lambda_2=0.2$, $\lambda_3=0.05$, and $n_1=0$, $n_2=n_3=1$, $N_1=N_2=1$, $N_3=N=0$.}}
\label{fig4}
\end{figure}

We will first consider what happens when  every $\mu_{k,l}^{coop}=0$ while $\mu_{k,l}^{ex}\neq0$. In this case the matrix $U$ is block-diagonal, and the computations are not particularly difficult. In Figures \ref{fig3} and \ref{fig4} we plot, as usual, $P_1(t)$, $P_2(t)$ and $P_3(t)$ as functions of time, for a particular choice of parameters and for two different initial conditions for $\Sc$: we fix $\mu_{1,2}^{ex}=0.2$, $\mu_{1,3}^{ex}=0.1$, $\mu_{2,3}^{ex}=0.15$, $\mu_{k,l}^{coop}=\tilde\lambda_j=0$, $\omega_1=0.1$, $\omega_2=\omega_3=0.2$, $
\Omega_1=\Omega_3=1$, $\Omega_2=2$, $\Omega=1$, $\lambda_1=0.1$, $\lambda_2=0.2$, $\lambda_3=0.05$. In Figure \ref{fig3} we  put  $n_1=0$, $n_2=n_3=1$, $N_1=0$, $N_2=N_3=N=1$, while in Figure \ref{fig4} we put  $n_1=0$, $n_2=n_3=1$, $N_1=0$, $N_2=N_3=1$ and $N=0$.

The difference with respect to the previous figures is evident: with this choice of parameters the parties reach a {\em final} decision, but there exists a certain time interval, the transient, during which the three decision functions oscillate between different possible choices. This result, in our opinion, reflects well what we observe in real life: many politicians say something one day and something different the day after. However, after some time, they really {\bf have} to decide, and this is described by the asymptotic values of our plots. Also the fact that $P_j(\infty)$'s are not really zero or one, but some intermediate value, reflects well the difficulty of taking a decision, so there is usually no sharp position, really. The only case when this happens, as we have already seen, is when  the parties interact just with their own reservoirs $\R_j$.

Incidentally, it might be useful to observe that, in order to see the asymptotic behavior of the functions $P_j(t)$, in Figures \ref{fig3} and \ref{fig4} we have used a larger time interval than that used in Figures \ref{fig1} and \ref{fig2}, since asymptotic limits become evident only in larger intervals (otherwise we would have seen just oscillations).

\subsection{No exchange effect}

\begin{figure}[ht]
\begin{center}
\includegraphics[width=0.4\textwidth]{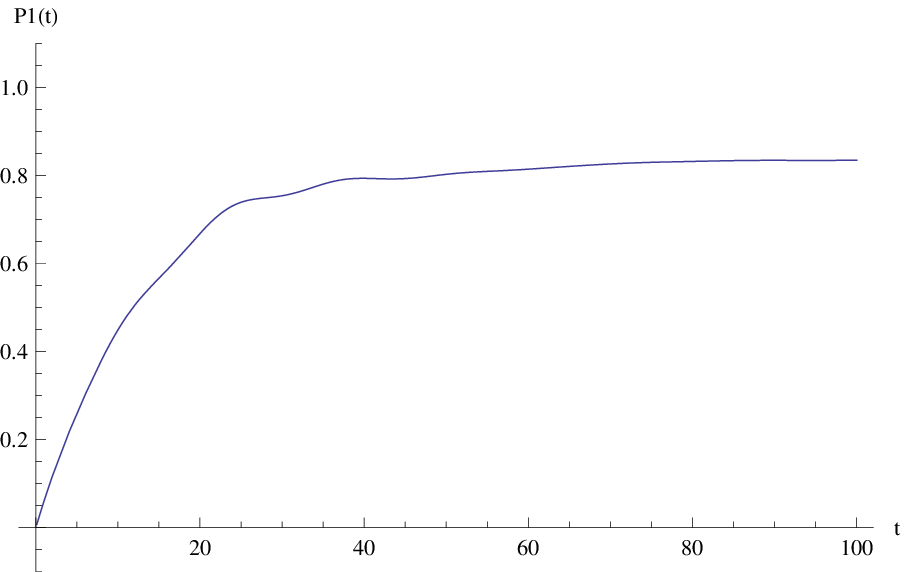}\hspace{8mm} %
\includegraphics[width=0.4\textwidth]{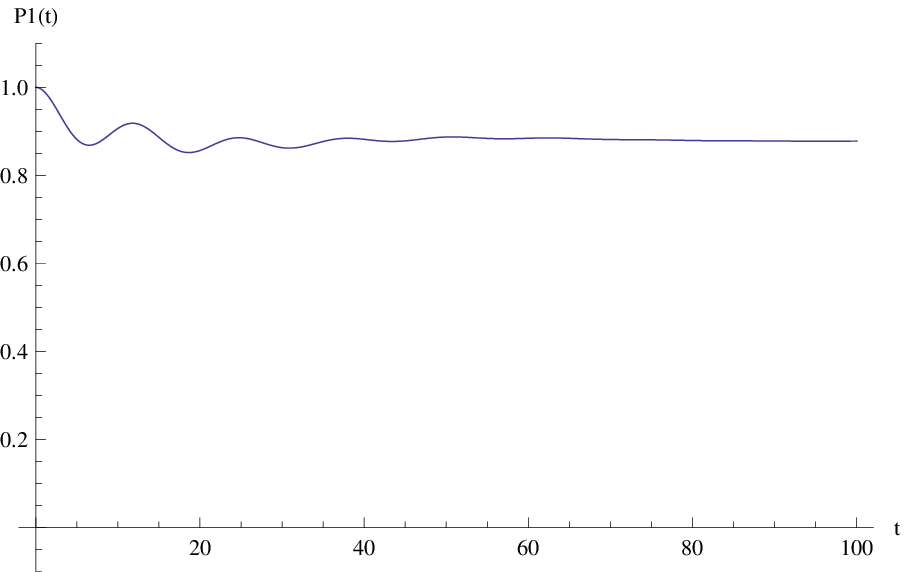}\hfill\\[0pt]
\includegraphics[width=0.4\textwidth]{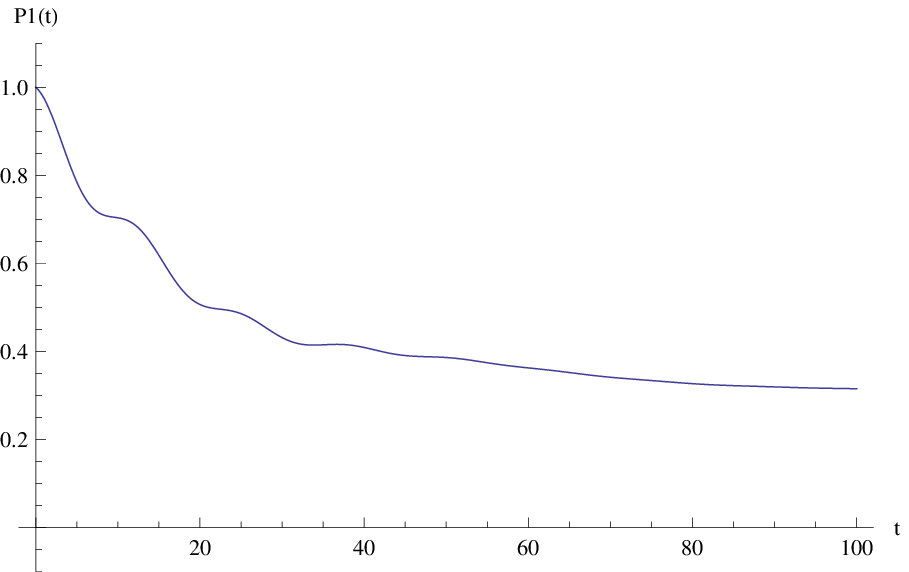}
\end{center}
\caption{{\protect\footnotesize $P_1(t)$ (top left), $P_2(t)$ (top right) and $P_3(t)$ (bottom) for $\mu_{1,2}^{coop}=0.1$, $\mu_{1,3}^{coop}=0.08$, $\mu_{2,3}^{coop}=0.1$, $\mu_{k,l}^{ex}=\tilde\lambda_j=0$, $\omega_1=0.1$, $\omega_2=\omega_3=0.2$, $
\Omega_1=\Omega_3=1$, $\Omega_2=2$, $\Omega=0.1$, $\lambda_1=0.1$, $\lambda_2=0.2$, $\lambda_3=0.05$, and $n_1=0$, $n_2=n_3=1$, $N_1=N_2=1$, $N_3=N=0$.}}
\label{fig5}
\end{figure}

\begin{figure}[ht]
\begin{center}
\includegraphics[width=0.4\textwidth]{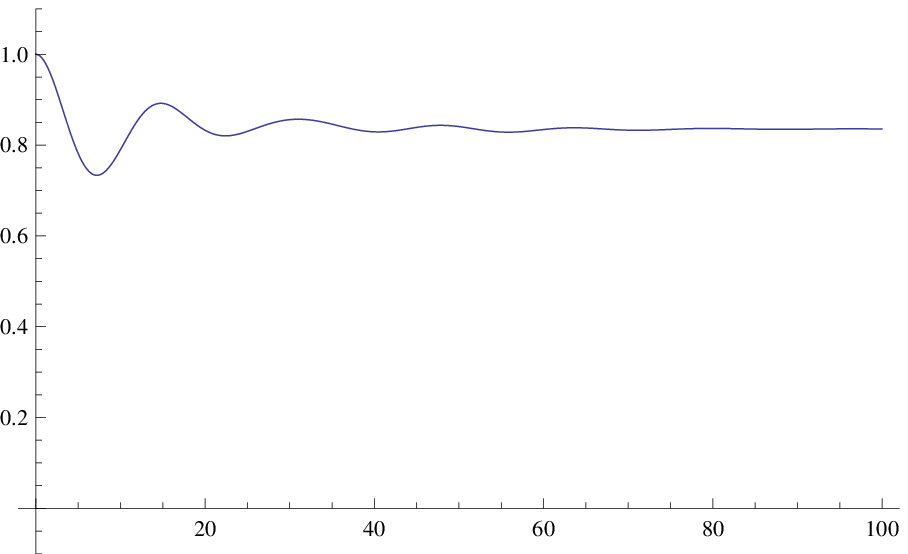}\hspace{8mm} %
\includegraphics[width=0.4\textwidth]{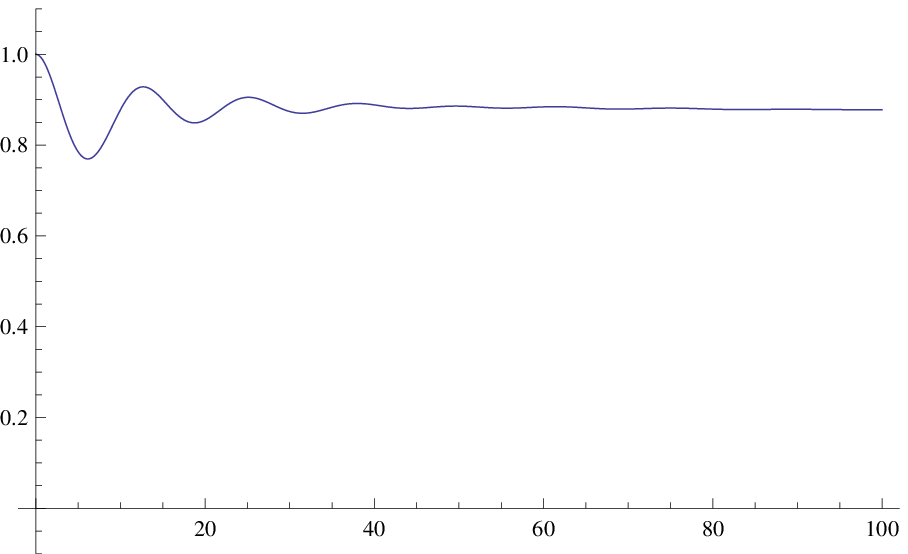}\hfill\\[0pt]
\includegraphics[width=0.4\textwidth]{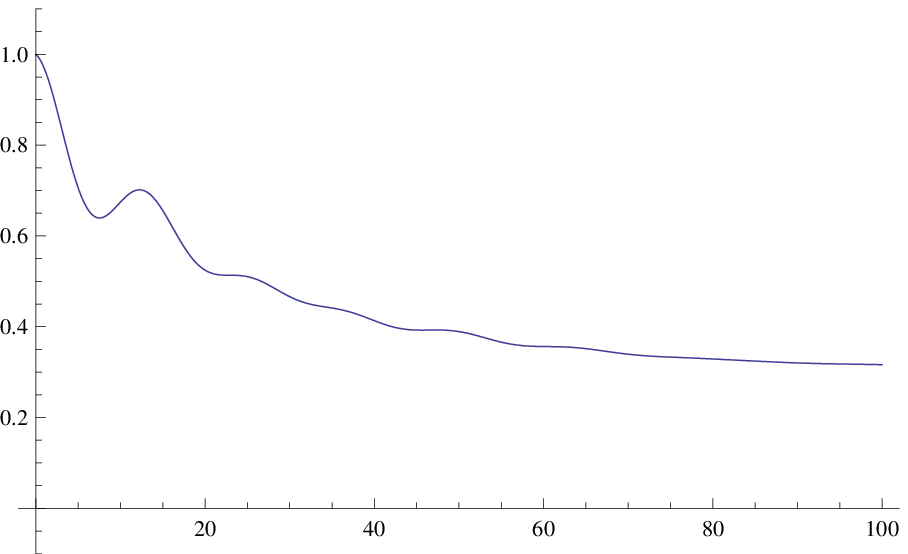}
\end{center}
\caption{{\protect\footnotesize $P_1(t)$ (top left), $P_2(t)$ (top right) and $P_3(t)$ (bottom) for $\mu_{1,2}^{coop}=0.1$, $\mu_{1,3}^{coop}=0.08$, $\mu_{2,3}^{coop}=0.1$, $\mu_{k,l}^{ex}=\tilde\lambda_j=0$, $\omega_1=0.1$, $\omega_2=\omega_3=0.2$, $
\Omega_1=\Omega_3=1$, $\Omega_2=2$, $\Omega=0.1$, $\lambda_1=0.1$, $\lambda_2=0.2$, $\lambda_3=0.05$, and $n_1=n_2=n_3=1$, $N_1=N_2=1$, $N_3=N=0$.}}
\label{fig6}
\end{figure}

Let us now put $\mu_{k,l}^{ex}=0$. In this case, since $\mu_{k,l}^{coop}\neq0$, $U$ has (almost) all non zero entries. Not unexpectedly, the computations are a bit harder. However, the plots we get do not differ much from those in Figures \ref{fig3} and \ref{fig4}. In fact, Figures \ref{fig5} and \ref{fig6} share with those ones the same main features, i.e. an initial oscillating behavior with a subsequent convergence to a certain asymptotic value. The parameters of Figures \ref{fig5} and \ref{fig6} coincide with those of Figures \ref{fig3} and \ref{fig4}, with the only difference that  here we put $\mu_{k,l}^{ex}=0$, whereas $\mu_{1,2}^{coop}=0.1$, $\mu_{1,3}^{coop}=0.08$, and $\mu_{2,3}^{coop}=0.1$. The initial conditions are given in the captions.

The conclusion is that the cooperative effect in the Hamiltonian produces almost the same effect as the exchange term in $H$. The only difference is that the oscillations look less evident, at least with the particular choice of the parameters considered here.

\subsection{Some further remarks}

The first important remark is that, if we start from one of the cases considered so far, and we slightly modify  the values of some of the parameters of the model (including also those which were previously chosen to be zero), the plots do not change much: the dynamics is stable under these changes. For instance, if we plot $P_j(t)$ fixing, as in Figure \ref{fig1}, $\mu_{k,l}^{ex}=0$, $\omega_1=1$, $\omega_2=\omega_3=2$, $
\Omega_1=\Omega_3=\Omega=0.1$, $\Omega_2=.2$, $\lambda_1=0.1$, $\lambda_2=0.2$, $\lambda_3=0.05$, $\tilde\lambda_1=0.1$,
$\tilde\lambda_2=0.2$, $\tilde\lambda_3=0$, and $n_1=0$, $n_2=n_3=1$, $N_1=N_2=1$, $N_3=N=0$ and letting $\mu_{k,l}^{coop}$ to be slightly greater than zero, the changes in the plots are extremely small, for sufficiently small $\mu_{k,l}^{coop}$'s. This, we believe, is due to the linearity of the equations of motion of the model.

\begin{figure}[ht]
\begin{center}
\includegraphics[width=0.4\textwidth]{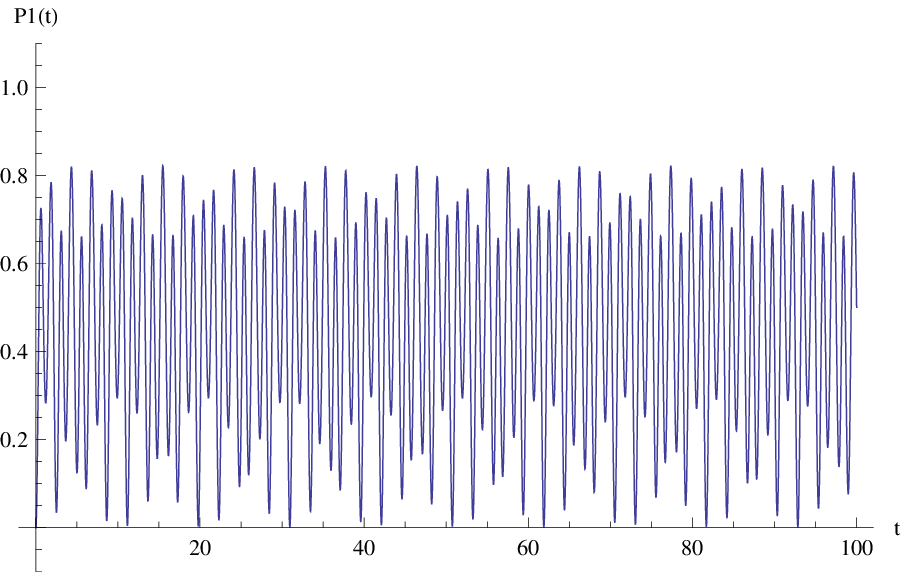}\hspace{8mm} %
\includegraphics[width=0.4\textwidth]{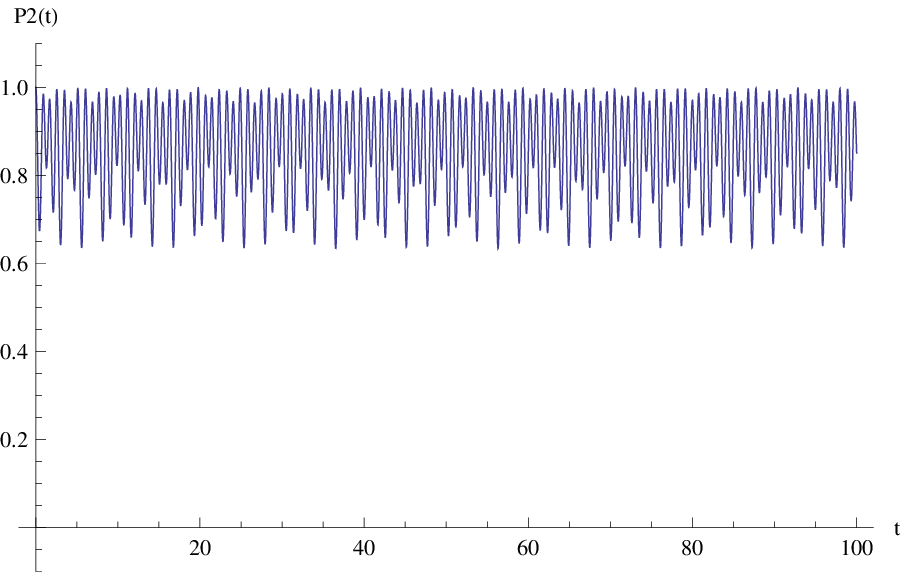}\hfill\\[0pt]
\includegraphics[width=0.4\textwidth]{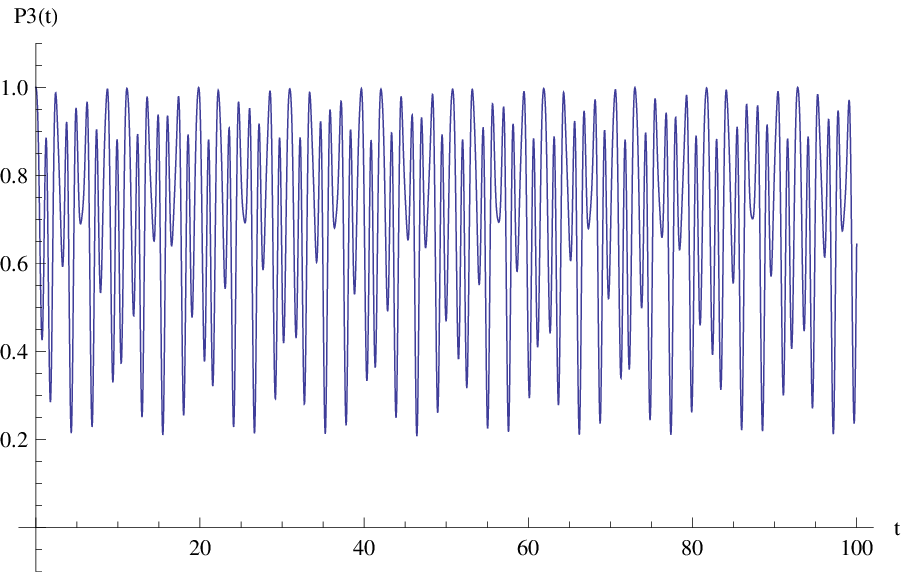}
\end{center}
\caption{{\protect\footnotesize $P_1(t)$ (top left), $P_2(t)$ (top right) and $P_3(t)$ (bottom) for $\mu_{1,2}^{coop}=0.1$, $\mu_{1,3}^{coop}=0.08$, $\mu_{2,3}^{coop}=0.1$, $\mu_{1,2}^{ex}=2$, $\mu_{1,3}^{ex}=1$, $\mu_{2,3}^{ex}=3$, $\tilde\lambda_j=\lambda_j=0$, $\omega_1=0.1$, $\omega_2=\omega_3=0.2$, $
\Omega_1=\Omega_3=\Omega=0.1$, $\Omega_2=0.2$ and $n_1=0$, $n_2=n_3=1$, $N_1=N_2=1$, $N_3=N=0$.}}
\label{fig7}
\end{figure}

More interesting for us is to see what happens when, for instance, $\mu_{k,l}^{ex}$ and $\mu_{k,l}^{coop}$ are different from zero, whereas $\lambda_j=\tilde\lambda_j=0$, $j=1,2,3$. The plots are given in Figure \ref{fig7}, for a particular choice of the parameters and the initial conditions. It is clear that there is no asymptotic value, at least at this time scale: on the contrary, the decision functions $P_j(t)$ appear to oscillate quite a bit in time. This can be easily understood: our choice on $\lambda_j$ and $\tilde\lambda_j$ is equivalent to the lack of any  backgrounds, and the parties only consult each other, exchanging what we could call, with a slight abuse of language, {\em quanta of decision} in a non conservative way (if $\mu_{k,l}^{coop}\neq0$). The parties have no input from their electors, and they are not able to decide what to do! Of course, as Figure \ref{fig7} shows, there are several maxima and minima in the decision functions corresponding to opposite attitudes. This is really observed in politics, where sometimes a decision changes with a very high frequency\footnote{In 2013, while Mr. Letta was the prime minister in Italy, the Members of the Parliament had to vote the {\em Fiducia}, i.e. they had to vote and say if Mr. Letta should resign or not. No more than ten minutes before the vote took place, an important exponent of PdL, Mr. Brunetta, said, in a TV interview, that PdL was \underline{not} going to vote in favor of Mr. Letta. However, only ten minutes after, Mr. Berlusconi, the President of the same party, talking to the other members of the House of Parliament announced they were going to vote in favor of Mr. Letta. This is exactly what we meant with {\em high frequency}!}.

Despite of the highly oscillating behavior shown in Figure \ref{fig7}, it is enough to add the effect of the reservoirs, by taking $\lambda_j$ or $\tilde\lambda_j$  larger than zero, to recover some asymptotic value (the smaller their values, the longer the time needed to reach this limit).

\vspace{2mm}

 In our analysis, we have also considered several other choices of parameters and initial conditions, including those in which all the parameters are non zero and not many differences have been found. The scheme which emerges is the following:

\begin{enumerate}

\item if there is no reservoir at all ($\lambda_j=\tilde\lambda_j=0$), the functions $P_j(t)$ oscillate, and no asymptotic value is reached. On the other hand, it is enough that $\lambda_j$ or $\tilde\lambda_j$ are slightly larger than zero, to recover some limiting value. This suggests that what really helps the various parties to get a decision is not the mutual interaction, but the interaction with the electors, and not necessarily those voting for them.

\item The exchange and the cooperative terms in the Hamiltonian produce a similar effect. The main difference is that, when $\mu_{k,l}^{coop}$ and the $\tilde\lambda_j$ are all zero, then an integral of motion exists, whereas this is not true if some of the $\mu_{k,l}^{coop}$ is non zero. Also, the amplitude of oscillations may be different.

\item The relative magnitude of the $\omega_j$'s (parameters of the parties) and of the $\Omega_j$'s (parameters of the electors) is important: in fact we have found a strong numerical evidence of the fact that when the $\Omega_j$'s are larger than the $\omega_j$'s, the plots oscillate much more than when the opposite happens or when they are of the same order of magnitude. In both cases, if $\lambda_j$ or $\tilde\lambda_j$ are non zero, $P_j(\infty)$ exists and it is a value between zero and one, but this value is essentially reached monotonically when $\omega_j\gg \Omega_k$, $j,k=1,2,3$, while it is reached after some (or many) oscillations when the opposite inequality holds. This result reflects similar conclusions deduced along the years and, in our understanding, it is related to the value of the parameters of the free part of the Hamiltonian of our models, \cite{bagbook}, which behaves, as already stated, as a sort of inertia.

\item Looking at the various plots, and to the analytic results, see (\ref{32}) and (\ref{33}) for instance, we see that an essential role is played by the reservoirs. The explicit values of the other parameters, however, may change the  asymptotic values of the various $P_j(\infty)$ and of the speed of convergence: the parties reach some stable decision, depending on whether they interact or not with the electors, but the decision, in general, is related not only to $N_j$ and $N$, but also to other parameters which, therefore, acquire an important role in the model. This is clear, for instance, in formula (\ref{31bis}), where we have fixed $N_1=1$ and $N=0$: if $\frac{\tilde\lambda_1^2\Omega_1}{\lambda_1^2\Omega}\simeq0$ then $P_1(\infty)\simeq1$, and $\Pc_1$ will try to form some coalition. On the other hand, if $\frac{\tilde\lambda_1^2\Omega_1}{\lambda_1^2\Omega}\gg0$, then $P_1(\infty)\simeq0$, and $\Pc_1$ will decide not to form any coalition.

\item Even if in some cases we have considered $P_j(\infty)$ as the {\em final decision } of $\Pc_j$, this does not mean that the parties decide how to behave only when $t$ is extremely large! This, of course, would be rather unpleasant. On the other hand all our plots show that an asymptotic value is reached, most of the times, sufficiently fast. This means that the decisions are made reasonably soon, and this procedure can be made even faster by changing the interaction parameters, as we have already discussed.

\end{enumerate}

\section{Conclusions}

In this paper we have modelled the interactions between three political parties, their electors and people who still have not decided who to vote for. Our main aim was to deduce the time behavior of what we have called {\em decision functions}, which describe the attitude of the various parties to give raise or not to some political coalition. The main result is that this decision can only be taken when the parties listen to the electors, since in this case $P_j(\infty)$ exists and is easily related to the attitude of the electors themselves. Otherwise, at least if some $\mu_{k,l}^{ex}$ or $\mu_{k,l}^{coop}$ is non zero, they simply oscillate between different attitudes towards alliances, making essentially no choice.

Concerning now other approaches to similar decision-making problems, we first consider the difference between our approach and the idea behind \cite{khren2}, which is also focused on the deduction of the time evolution of some decision functions. Our feeling is that our approach has a more direct interpretation, since we use an Hamiltonian where all terms have a clear meaning. On the other hand, the master equation used in \cite{khren2}, makes less evident any comparison with the real system. Moreover, using approaches as in \cite{khren2}, the role of the electors would be somehow hidden because there is no explicit reservoir at all. But this is, probably, just a matter of personal taste. More on the line of what is done in this paper can be found in \cite{marti} and \cite{buse2}, where the dynamics of the system is produced by some very simple Hamiltonians adopting an Heisenberg-like approach, as we do in Section II. The main difference, we believe, is that while our Hamiltonian describes in some details several {\em realistic} interactions, the ones adopted in \cite{marti} and \cite{buse2} are so simple that they can only be considered as {\em effective Hamiltonians}, surely useful to describe some special aspect of the physical system, but constructed rather ad-hoc.

Of course, several things can still be done in our approach: first, a more detailed analysis of the role of the parameters could be interesting. But, more interesting to us, once we have deduced that one party wants to form a coalition, one could wonder: with which other party? We believe that this question can still be answered using our strategy, and this is part of our future work.

\section*{Acknowledgements}

The author acknowledges partial support from Palermo University and from G.N.F.M. of the INdAM. The author also thanks the referees for their quite useful suggestions, and Dr. Giorgia Bellomonte for her careful reading of the manuscript.

\renewcommand{\theequation}{A.\arabic{equation}}

\section*{Appendix:  Few results on the number representation}
To keep the paper self-contained, we repeat what already done in previous papers, \cite{bagoliv1,bagoliv2}, and we introduce here few important facts in quantum mechanics and in the so--called number representation.  More details can be found, for instance, in \cite{mer,rom}, as well as in\cite{bagbook}.

Let $\Hil$ be an Hilbert space, and $B(\Hil)$ the set of all (bounded) operators on $\Hil$.    Let $\ST$ be our physical system, and $\A$ the
set of all those operators useful for a complete description of $\ST$, which includes the \emph{observables} of $\ST$. For simplicity, it is
convenient (but not really necessary) to assume that  $\A$ coincides with $B(\Hil)$ itself. The description of the time evolution of $\ST$ is related to a self--adjoint
operator $H=H^\dagger$ which is called the \emph{Hamiltonian} of $\ST$, and which in standard quantum mechanics represents  the energy of
$\ST$. In this paper we have adopted the so--called \emph{Heisenberg} representation, in which the time evolution of an observable $X\in\A$ is given by
\be X(t)=\exp(iHt)X\exp(-iHt), \label{a1} \en or, equivalently, by the solution of the differential equation \be
\frac{dX(t)}{dt}=i\exp(iHt)[H,X]\exp(-iHt)=i[H,X(t)],\label{a2} \en where $[A,B]:=AB-BA$ is the \emph{commutator} between $A$ and $B$. The time
evolution defined in this way is a one--parameter group of automorphisms of $\A$.

An operator $Z\in\A$ is a \emph{constant of motion} if it commutes with $H$. Indeed, in this case, equation (\ref{a2}) implies that $\dot
Z(t)=0$, so that $Z(t)=Z$ for all $t$.

In some previous applications, \cite{bagbook}, a special role was played by the so--called \emph{canonical commutation
relations}. Here, these are replaced by the so--called \emph{canonical anti--commutation relations} (CAR): we say that a set of operators
$\{a_\ell,\,a_\ell^\dagger, \ell=1,2,\ldots,L\}$ satisfies the CAR if the conditions \be \{a_\ell,a_n^\dagger\}=\delta_{\ell n}\1,\hspace{8mm}
\{a_\ell,a_n\}=\{a_\ell^\dagger,a_n^\dagger\}=0 \label{a3} \en hold true for all $\ell,n=1,2,\ldots,L$. Here, $\1$ is the identity operator
and $\{x,y\}:=xy+yx$ is the {\em anticommutator} of $x$ and $y$. These operators, which are widely analyzed in any textbook about quantum
mechanics (see,  for instance, \cite{mer,rom}) are those which are used to describe $L$ different \emph{modes} of fermions. From these
operators we can construct $\hat n_\ell=a_\ell^\dagger a_\ell$ and $\hat N=\sum_{\ell=1}^L \hat n_\ell$, which are both self--adjoint. In
particular, $\hat n_\ell$ is the \emph{number operator} for the $\ell$--th mode, while $\hat N$ is the \emph{number operator of $\ST$}.
Compared with bosonic operators, the operators introduced here satisfy a very important feature: if we try to square them (or to rise them to higher
powers), we simply get zero: for instance, from (\ref{a3}), we have $a_{\ell}^2=0$. This is related to the fact that fermions satisfy the Fermi
exclusion principle \cite{rom}.

The Hilbert space of our system is constructed as follows: we introduce the \emph{vacuum} of the theory, that is a non zero vector $\varphi_{\bf 0}$
which is annihilated by all the operators $a_\ell$: $a_\ell\varphi_{\bf 0}=0$ for all $\ell=1,2,\ldots,L$. Such a  vector surely exists. Then we act on $\varphi_{\bf 0}$
with the  operators $a_\ell^\dagger$ (but not with higher powers, since these powers are simply zero!): \be
\varphi_{n_1,n_2,\ldots,n_L}:=(a_1^\dagger)^{n_1}(a_2^\dagger)^{n_2}\cdots (a_L^\dagger)^{n_L}\varphi_{\bf 0}, \label{a4} \en $n_\ell=0,1$ for
all $\ell$. These vectors form an orthonormal set and are eigenstates of both $\hat n_\ell$ and $\hat N$: $\hat
n_\ell\varphi_{n_1,n_2,\ldots,n_L}=n_\ell\varphi_{n_1,n_2,\ldots,n_L}$ and $\hat N\varphi_{n_1,n_2,\ldots,n_L}=N\varphi_{n_1,n_2,\ldots,n_L},$
where $N=\sum_{\ell=1}^Ln_\ell$. Moreover, using CAR's, we deduce that $$\hat
n_\ell\left(a_\ell\varphi_{n_1,n_2,\ldots,n_L}\right)=(n_\ell-1)(a_\ell\varphi_{n_1,n_2,\ldots,n_L})$$ and $$\hat
n_\ell\left(a_\ell^\dagger\varphi_{n_1,n_2,\ldots,n_L}\right)=(n_\ell+1)(a_l^\dagger\varphi_{n_1,n_2,\ldots,n_L}),$$ for all $\ell$. Then
 $a_\ell$ and $a_\ell^\dagger$ are  called the
\emph{annihilation} and the \emph{creation} operators. Notice that, in some sense, $a_\ell^\dagger$ is {\bf also} an annihilation operator since,
acting on a state with $n_\ell=1$, it destroys that state.

The Hilbert space $\Hil$ is obtained by taking  the linear span of all these vectors. Of course, $\Hil$ has a finite dimension. In particular,
for just one mode of fermions, $dim(\Hil)=2$. This also implies that, contrarily to what happens for bosons, all the fermionic operators are
bounded.

The vector $\varphi_{n_1,n_2,\ldots,n_L}$ in (\ref{a4}) defines a \emph{vector (or number) state } over the algebra $\A$  as \be
\omega_{n_1,n_2,\ldots,n_L}(X)= \langle\varphi_{n_1,n_2,\ldots,n_L},X\varphi_{n_1,n_2,\ldots,n_L}\rangle, \label{a5} \en where
$\langle\,,\,\rangle$ is the scalar product in  $\Hil$. As we have discussed in \cite{bagbook}, these states are useful
to \emph{project} from quantum to classical dynamics and to fix the initial conditions, as we have done in (\ref{add1}).

\end{document}